\documentclass[amsmath,amssymb,aps,prb,twocolumn, superscriptaddress,longbibliography]{revtex4-1}
\usepackage{amsmath}
\usepackage{hyperref}
\usepackage{color}
\usepackage{graphicx}

\begin{document}
\newcommand{\p}{\partial}
\newcommand{\eps}{\varepsilon}
\newcommand{\om}{\omega}
\newcommand{\red}[1]{{\color{red}#1}}
\newcommand{\blue}[1]{{\color{blue}#1}}
\newcommand{\ds}[2]{\frac{\partial^2 #1}{\partial #2^2}}
\newcommand{\pol}[1]{\text{polylog}\,\left(#1\right)}

\title{Nonlocal response of Mie-resonant dielectric particles}

\author{Daniel A. Bobylev}
\affiliation{Physics and Engineering Department, ITMO University, Saint Petersburg 197101, Russia}

\author{Daria A. Smirnova}
\affiliation{Nonlinear Physics Centre, Australian National University, Canberra ACT 2601, Australia}
\affiliation{Institute of Applied Physics, Russian Academy of Science, Nizhny Novgorod 603950, Russia}

\author{Maxim A. Gorlach}
\affiliation{Physics and Engineering Department, ITMO University, Saint Petersburg 197101, Russia}

% \date{\today}

\begin{abstract}
Mie-resonant high-index dielectric particles are at the core of modern all-dielectric photonics. In many situations, their response to the external fields is well-captured by the dipole model which neglects the excitation of higher-order multipoles. In that case, it is commonly assumed that the dipole moments induced by the external fields are given by the product of particle polarizability tensor and the field in the particle center. Here, we demonstrate that the dipole response of non-spherical subwavelength dielectric particles is significantly more complex since the dipole moments are defined not only by the field in the particle center but also by the second-order spatial derivatives of the field. As we prove, such nonlocal response is especially pronounced in the vicinity of anapole minimum in the scattering cross-section. We examine the excitation of high-index dielectric disk in microwave domain and silicon nanodisk in near infrared applying group-theoretical analysis and retrieving the nonlocal corrections to the dipole moments. Extending the discrete dipole model to include nonlocality of the dipole response, we demonstrate an improved agreement with full-wave numerical simulations. These results provide important insights into meta-optics of Mie-resonant non-spherical particles as well as metamaterials and metadevices based on them.
\end{abstract}

\maketitle

\section{Introduction}

Over the recent years, all-dielectric nanophotonics and meta-optics~\cite{Kuznetsov-Science,Kruk-Kivshar} have demonstrated a variety of exciting functionalities including strong directional scattering of light~\cite{Staude-13,Decker}, flexible phase manipulation of the transmitted signal with transparent metasurfaces~\cite{Kruk-APL}, high-quality modes of dielectric particles~\cite{Koshelev}, enhanced nonlinear phenomena in the arrays of resonant nanoparticles~\cite{Shcherbakov,Koshelev-Science} and precise molecular fingerprinting with all-dielectric metasurfaces~\cite{Tittl}. The physics underlying this plethora of effects is based on the combination of electric and magnetic responses of dielectric particles~\cite{Evlyukhin-NL,Kuznetsov-SciRep,Smirnova-Optica}, where  optical magnetic response is associated with circular displacement currents excited in the scatterer.

A key theoretical tool to capture electromagnetic behavior of subwavelength objects is multipole expansion~\cite{Jackson} which presents the field scattered by the particle as a sum of different multipoles, each of them being characterized by the unique set of polarization and angular dependence of the radiation pattern. In many situations, the dominant contribution to the scattering cross-section of subwavelength particles is provided by the lowest-order multipoles, namely, magnetic and electric dipoles, while the contribution of higher-order multipoles can be neglected. 

In that case, the physics of complex particle arrays can be efficiently explored using the discrete dipole model~\cite{Purcell,Draine,Yurkin,Evlyukhin-11}. In this approach, the scatterer is viewed as electric and magnetic dipoles placed in its center, while all information on particle properties is embedded into its electric and magnetic polarizability tensors. These tensors link the external fields acting on the particle to its dipole moments and can be retrieved from full-wave numerical simulations. 

However, dielectric particles used in experimentally relevant situations are not that deeply subwavelength, and the diameter of the particle has typically the same order of magnitude as the wavelength of light at magnetic or electric dipole resonance~\cite{Kruk-Kivshar}. Therefore, it is not obvious {\it a priori} that electric and magnetic dipole moments of an arbitrarily shaped particle are related only to the field in the particle center.

In this Article, we assess this important assumption and reveal that the dipole moments of the typical Mie-resonant disk are governed not only by electric and magnetic fields in its center but also by their second-order spatial derivatives which crucially determine the electromagnetic response of the disk in the vicinity of anapole minimum~\cite{Baryshnikova-adom} in the scattering cross-section. To highlight that the discussed physics is universally valid across the entire electromagnetic spectrum, we examine two representative examples: dielectric disk made of high-permittivity ceramics with resonances in the microwave domain and silicon nanodisk supporting Mie resonances in the near infrared. In both cases, we focus on the frequency range where electric and magnetic dipole responses provide the dominant contribution to the scattering cross-section separating them from the contributions of higher-order multipoles via multipole decomposition technique.

Furthermore, we extend the discrete dipole model by incorporating the dependence of the dipole moments on spatial derivatives of the fields and demonstrate an improved accuracy of such approach compared to the conventional discrete dipole approximation.

It should be stressed that our results are conceptually different from the nonlocal effects in small metallic nanoparticles manifested via size-dependent resonance shifts and linewidth broadening~\cite{Raza,Abajo}. The latter effects occurring due to electron-electron interactions are enhanced as the size of plasmonic nanoparticle decreases. In stark contrast, the mechanism we discuss here becomes increasingly important as the size of the particle grows becoming of the same order of magnitude as the wavelength at the dipole resonance.

The rest of the article is organized as follows. In Section~\ref{sec:Symmetry} we construct a general relation between the dipole moment of the particle from one side and incident field with its spatial derivatives from the other using the symmetry arguments. To support our analysis further, in Sec.~\ref{sec:Simulations} we consider a dielectric disk made of high-permittivity ceramics in the microwave domain and a nanodisk made of crystalline silicon in the near infrared. The parameters of both particles are chosen in such a way that the dipole response dominates the rest of multipole contributions in a sufficiently wide frequency range. Performing full-wave numerical simulations of the disks' dipole response, we reveal the conditions for the strongest nonlocal effects. Focusing on the case of ceramic disk, we extract the components of polarizability tensors as well as the nonlocal corrections to the induced dipole moments. Section~\ref{sec:Meta} continues with the discussion of the generalized discrete-dipole model which incorporates nonlocal corrections to the dipole moments and enables an increased accuracy in the description of meta-crystals composed of such Mie-resonant disks. Finally, Sec.~\ref{sec:Concl} concludes with a summary and an outlook for future studies.

\section{Symmetry analysis of the disk response}\label{sec:Symmetry}

In the most general scenario, electric and magnetic dipole moments ${\bf d}$ and ${\bf m}$ of the particle induced by the impinging plane wave can be presented as Taylor series with respect to wave vector ${\bf k}$:
\begin{gather}
d_i = \eps_0\,\left\lbrace\alpha_{ij}\,E_j + \beta_{ijl}\,E_j k_l + \gamma_{ijlm} \,E_j k_l k_m + \dots\right\rbrace\:, \label{eq1}\\
m_i=\alpha_{ij}^{(\rm{m})}\,H_j + \beta_{ijl}^{(\rm{m})}\,H_j k_l + \gamma_{ijlm}^{(\rm{m})} H_j k_l k_m + \dots\:,\label{eq1b}
\end{gather}
where SI system of units and $e^{-i\om\,t}$ time convention are used. $n^{\rm{th}}$ terms of both expansions scale as $(R/\lambda)^n$ relative to the respective leading-order terms, where $R$ is the particle characteristic size and $\lambda$ is the wavelength. Therefore, assuming that the particle is subwavelength, we keep only the first three terms in Eqs.~\eqref{eq1}, \eqref{eq1b}.

Note that in the analysis below we define the dipole moment based on the angular dependence of the fields, consistently with Refs.~\cite{Rockstuhl-OE-15,Rockstuhl-OC-18}. In the alternative formulations of multipole expansion~\cite{Evlyukhin-16,Gurvitz}, however, thus defined $d_i$ and $m_i$ correspond to the sum of dipole moment, toroidal moment and higher-frequency contributions.

Tensors $\beta_{ijl}$ and $\beta_{ijl}^{(\rm{m})}$ describe bianisotropic response of the particle being zero for any inversion-symmetric configuration including the case of the disk. Furthermore, $D_{\infty h}$ symmetry group of the disk ensures that the tensors $\alpha_{ij}$, $\alpha_{ij}^{(\rm{m})}$ and $\gamma_{ijlm}$, $\gamma_{ijlm}^{(\rm{m})}$ have two and six independent components, respectively, which strongly simplifies the analysis (see the details in Appendix~\ref{app:symmetry}). Due to symmetry, these tensors can be constructed only from Kronecker symbols $\delta_{ij}$ and even powers of ${\bf n}$ vector directed along the disk axis. Besides that, $\alpha_{ij}$ and $\gamma_{ijlm}$ are symmetric with respect to the first pair $(i,j)$ of the indices due to symmetry of kinetic coefficients~\cite{landau5}; $\gamma_{ijlm}$ is also symmetric with respect to the last pair $(l,m)$ of the indices. The above requirements yield:
\begin{gather}
\alpha_{ij} = \alpha_1\, \delta_{ij} + \alpha_2\, n_i n_j\:,\label{Alpha}\\
\gamma_{ijlm} = \gamma_1\, \delta_{ij}\, n_l n_m +\gamma_2\,\left(n_i n_l\, \delta_{jm} + n_j n_m \,\delta_{il}\right. \notag\\
\left.+n_j n_l\,\delta_{im} + n_i n_m\,\delta_{jl}\right) + \gamma_3\, n_i n_j n_l n_m\label{Gamma}\\
+\gamma_4\,\delta_{ij} \delta_{lm} + \gamma_5\,\left(\delta_{il} \delta_{jm} + \delta_{im} \delta_{jl}\right)
+\gamma_6\, n_i n_j \delta_{lm}\:,\notag
\end{gather}
where $\alpha_{1,2}$ and $\gamma_{1-6}$ are some unknown scalar coefficients which depend on the material and shape of the particle and the frequency of excitation.

Combining Eqs.~\eqref{Alpha}, \eqref{Gamma} with Eqs.~\eqref{eq1}, \eqref{eq1b}, we derive the expression for the dipole moment of the disk:
\begin{gather}
d_i = \alpha_{\perp} \eps_0\,E_i + (\alpha_{\parallel} - \alpha_{\perp}) n_i \,\eps_0\,E_z + \gamma_1 \,\eps_0\,E_i k_z^2 \notag\\
+2\gamma_2 k_i k_z\,\eps_0\,E_z  + \gamma_3 n_i \,\eps_0\,E_z k_z^2,\label{dipole}
\end{gather}
where we take into account that $|{\bf k}|=k_0=\om/c$ and ${\bf k}\cdot{\bf E}=0$ since the incident field satisfies the condition $\text{div}\,{\bf E}=0$. $\alpha_{\perp}$ and $\alpha_{\parallel}$ are the standard frequency-dependent components of polarizability tensor of an anisotropic particle defined as $ \alpha_{\perp} = \alpha_1 + \gamma_4\,k_0^2 $ and $\alpha_{\parallel} = \alpha_{\perp} + \alpha_2 + \gamma_6\,k_0^2$. Similar equation holds for the magnetic dipole moment:
\begin{gather}
m_i = \alpha_{\perp}^{(\rm{m})} H_i + (\alpha_{\parallel}^{(\rm{m})} - \alpha_{\perp}^{(\rm{m})}) n_i H_z + \gamma_1^{(\rm{m})} H_i k_z^2 \notag\\
+2\gamma_2^{(\rm{m})} k_i k_z\, H_z  + \gamma_3^{(\rm{m})} n_i H_z k_z^2\:.\label{mdipole}
\end{gather}

Quite importantly, if the particle is spherical and has full rotational symmetry, the only nonzero components of the tensors Eqs.~\eqref{Alpha}, \eqref{Gamma} are $\alpha_1$, $\gamma_4$ and $\gamma_5$, which ensures that the link between the dipole moment and the field remains local.

\begin{figure}[t]
	\includegraphics{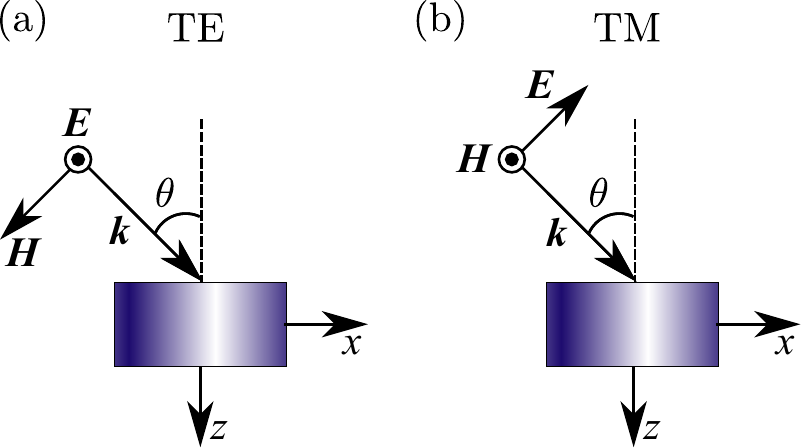}
	\caption{Excitation of the disk by the incident plane wave. (a)~TE-polarized excitation which induces $d_y$, $m_x$ and $m_z$ components of dipole moments; (b)~TM-polarized excitation which induces $d_x$, $d_z$ and $m_y$ components of dipole moments.}
	\label{fig:disk}
\end{figure}

In the case of a disk, second-order nonlocal corrections to the dipole moment of the particle are captured by the three additional terms proportional to $\gamma_1$, $\gamma_2$ and $\gamma_3$. Note that all of them exhibit characteristic dependence on the direction of the incident wave propagation since they depend on $k_z$. Based on this observation, we consider the geometry illustrated in Fig.~\ref{fig:disk}.

TE-polarized wave [Fig.~\ref{fig:disk}(a)] excites $d_y$, $m_x$ and $m_z$ components of the dipole moments given by the equations
\begin{gather}
d_y=\left(\alpha_{\bot}+\gamma_1 k_0^2 \cos^2\theta\right)\,\eps_0\,E_0\:,\label{dy}\\
m_x = -\left\lbrace\left(\alpha_{\perp}^{(\rm{m})} - 2\gamma_2^{(\rm{m})} k_0^2\right)\,\cos\theta\right. \notag\\
\left.+\left(\gamma_1^{(\rm{m})} + 2\gamma_2^{(\rm{m})}\right)\, k_0^2 \cos^3\theta\right\rbrace\, H_0\:,\label{mx}\\
m_z = \left\lbrace \left[\alpha_{\parallel}^{(\rm{m})} + \left(\gamma_1^{(\rm{m})} + 2\gamma_2^{(\rm{m})} + \gamma_3^{(\rm{m})}\right) k_0^2\right] \sin\theta\right. \notag\\
\left.-\left(\gamma_1^{(\rm{m})} + 2\gamma_2^{(\rm{m})} + \gamma_3^{(\rm{m})}\right)\, k_0^2 \sin^3\theta\right\rbrace H_0\:,\label{mz}
\end{gather}
whereas TM-polarized excitation [Fig.~\ref{fig:disk}(b)] results in $d_x$, $d_z$ and $m_y$ components of the dipole moments:
\begin{gather}
m_y=\left(\alpha^{(\rm{m})}_{\bot}+\gamma_1^{(\rm{m})}\,k_0^2 \cos^2\theta\right)H_0\:,\label{my}\\
d_x = \left\lbrace\left(\alpha_{\perp} - 2\gamma_2 k_0^2\right) \cos\theta\right.\notag \\
\left.+\left(\gamma_1 + 2\gamma_2\right)\,k_0^2 \cos^3\theta\right\rbrace\, \eps_0\,E_0\:,\label{dx}\\
d_z = \left\lbrace-\left[\alpha_{\parallel} + \left(\gamma_1 + 2\gamma_2 + \gamma_3\right)\,k_0^2\right] \sin\theta\right. \notag\\
\left.+\left(\gamma_1 + 2\gamma_2 + \gamma_3\right) k_0^2 \sin^3\theta\right\rbrace\, \eps_0\,E_0\:.\label{dz}
\end{gather}
Hence, all relevant coefficients can be extracted by fitting the dependence of dipole moments on the incidence angle $\theta$ of the plane wave.

It should be stressed that the symmetry analysis above is fully general and not limited only to the propagating fields. To capture the nonlocal response of the disk to the arbitrary excitation, we replace $k_n$ by $-i\,\partial_n$ in Eqs.~\eqref{dipole}-\eqref{mdipole}, which yields:
\begin{gather}
d_i = \alpha_{\perp} \eps_0 E_i + (\alpha_{\parallel} - \alpha_{\perp}) n_i \eps_0 E_z - \gamma_1 \eps_0 \frac{\partial^2 E_i}{\partial z^2}\notag\\
-2\gamma_2 \eps_0 \frac{\partial^2 E_z}{\partial x_i \partial z} - \gamma_3 n_i \eps_0 \frac{\partial^2 E_z}{\partial z^2}\:,\label{DipoleFull}\\
m_i = \alpha_{\perp}^{(\rm{m})} H_i + (\alpha_{\parallel}^{(\rm{m})} - \alpha_{\perp}^{(\rm{m})}) n_i H_z - \gamma_1^{(\rm{m})} \frac{\partial^2 H_i}{\partial z^2}\notag\\
-2\gamma_2^{(\rm{m})} \frac{\partial^2 H_z}{\partial x_i \partial z}- \gamma_3^{(\rm{m})} n_i \frac{\partial^2 H_z}{\partial z^2}\:.\label{MdipoleFull}
\end{gather}

\section{Numerical simulations of the nonlocal response}\label{sec:Simulations}
 
\begin{figure}[hb]
	\includegraphics[width=0.9\linewidth]{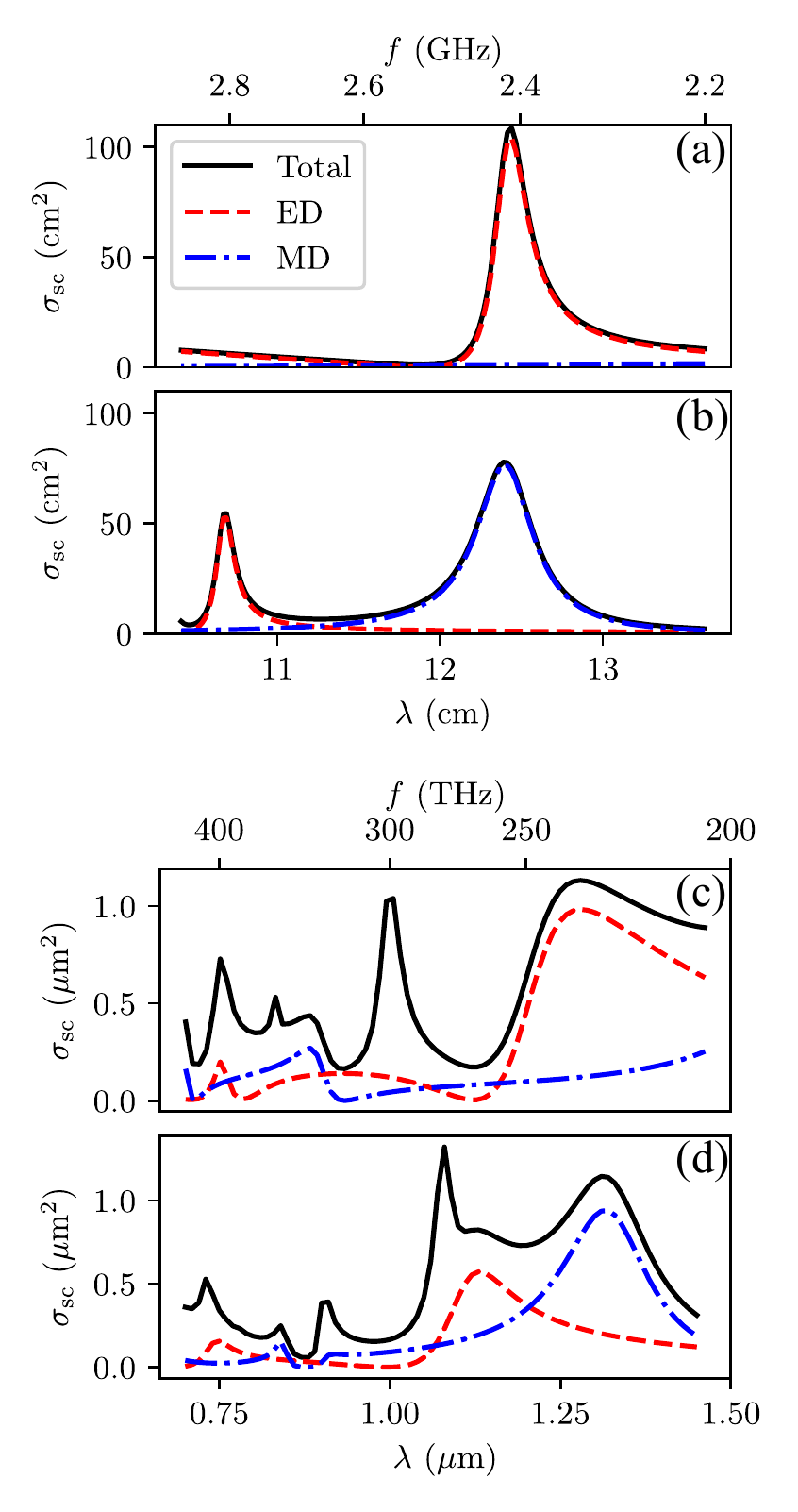}
	\caption{Scattering spectra of the disk illuminated by the plane wave propagating along $x$ axis in geometry of Fig.~\ref{fig:disk}. Solid line shows the total scattering cross-section, red dashed and blue dot-dashed curves show the contribution of electric and magnetic dipoles, respectively. (a,b) Results for ceramic disk with permittivity $\varepsilon = 39+0.078\,i$, radius $R = 14.55$~mm and height $h = 11.61$~mm. (c,d) Results for nanodisk made of crystalline silicon with radius $R=242$~nm and height $h=220$~nm. (a,c) TE-polarized plane wave excitation with electric field along $y$ axis. (b,d) TM-polarized plane wave excitation with electric field along $z$ axis.}
\label{fig:spectrum}
\end{figure}  
 
As experimentally relevant examples of cylindrical particles we consider two cases. The first one corresponds to the disk made of high-permittivity ceramics $\varepsilon = 39 + 0.078\,i$ with radius $R = 14.55$~mm and height $h = 11.61$~mm close to the values used in the recent experiments~\cite{GorlachPRB2019}. Chosen parameters ensure that electric and magnetic dipole resonances residing in the range $(2.2 \div 2.9)$~GHz are well-separated from the higher-order multipole resonances. Furthermore, $2R/\lambda\approx 0.24$ in the chosen frequency range, which means that the size of the disk is of the same order of magnitude as wavelength.

Examining the response of the disk to the incident TE-polarized plane wave [geometry Fig.~\ref{fig:disk}(a), $\theta=\pi/2$], we recover a single characteristic peak in the scattering cross-section at frequencies around 2.4~GHz [Fig.~\ref{fig:spectrum}(a)]. Multipole analysis of the scattered field reveals the dominant contribution of   electric dipole, while the contributions from magnetic dipole and from higher-order multipoles are strongly suppressed. This means that the incident field excites in-plane electric dipole resonance of the disk. Note also that the scattering spectrum has a pronounced minimum at frequencies around $2.5$~GHz, which corresponds to the so-called anapole.

TM-polarized excitation [geometry Fig.~\ref{fig:disk}(b), $\theta=\pi/2$] gives rise to the two scattering peaks [Fig.~\ref{fig:spectrum}(b)]: one around 2.4~GHz with the dominant contribution of in-plane magnetic dipole and another one around 2.8~GHz corresponding to $z$-oriented electric dipole. The contribution of higher-order multipoles to the scattering cross-section in the frequency range $(2.20\div 2.75)$~GHz is below 4~cm$^2$ and 0.25~cm$^2$ for TE- and TM-polarized excitations, respectively. Hence, the dipole model is clearly adequate in this case.

As a second example of a cylindrical scatterer, we consider a nanodisk made of crystalline silicon with radius $R=242$~nm and height $h=220$~nm as in the recent experiments~\cite{Decker}. Such disk supports dipole resonances in the near infrared spectral range: $(200 \div 275)$~THz so that the particle size is also comparable to wavelength: $2R/\lambda\approx 0.38$. Since the refractive index of silicon is lower than that of microwave ceramics, the relative spectral separation of Mie resonances is smaller. Nevertheless, the dipole response of the disk dominates at wavelengths $1.2<\lambda<1.5$~$\mu$m featuring the same structure of the scattering peaks as its microwave counterpart [Fig.~\ref{fig:spectrum}(c,d)].

To quantify the nonlocal dipole response of both scatterers, we examine their excitation by TE-polarized plane wave for the two incidence angles: $\theta=0$ and $\theta=\pi/2$. Naively, one would expect that the dipole moments $d_y(0)$ and $d_y(\pi/2)$ induced by the impinging wave should be the same since $E_y$ component of the incident field does not depend on the incidence angle. Therefore, the difference $|d_y(\pi/2)-d_y(0)|$ provides a direct measure of nonlocality [cf. Eq.~\eqref{dy}].

\begin{widetext}

    \begin{figure}[h]
	\includegraphics[width=1\linewidth]{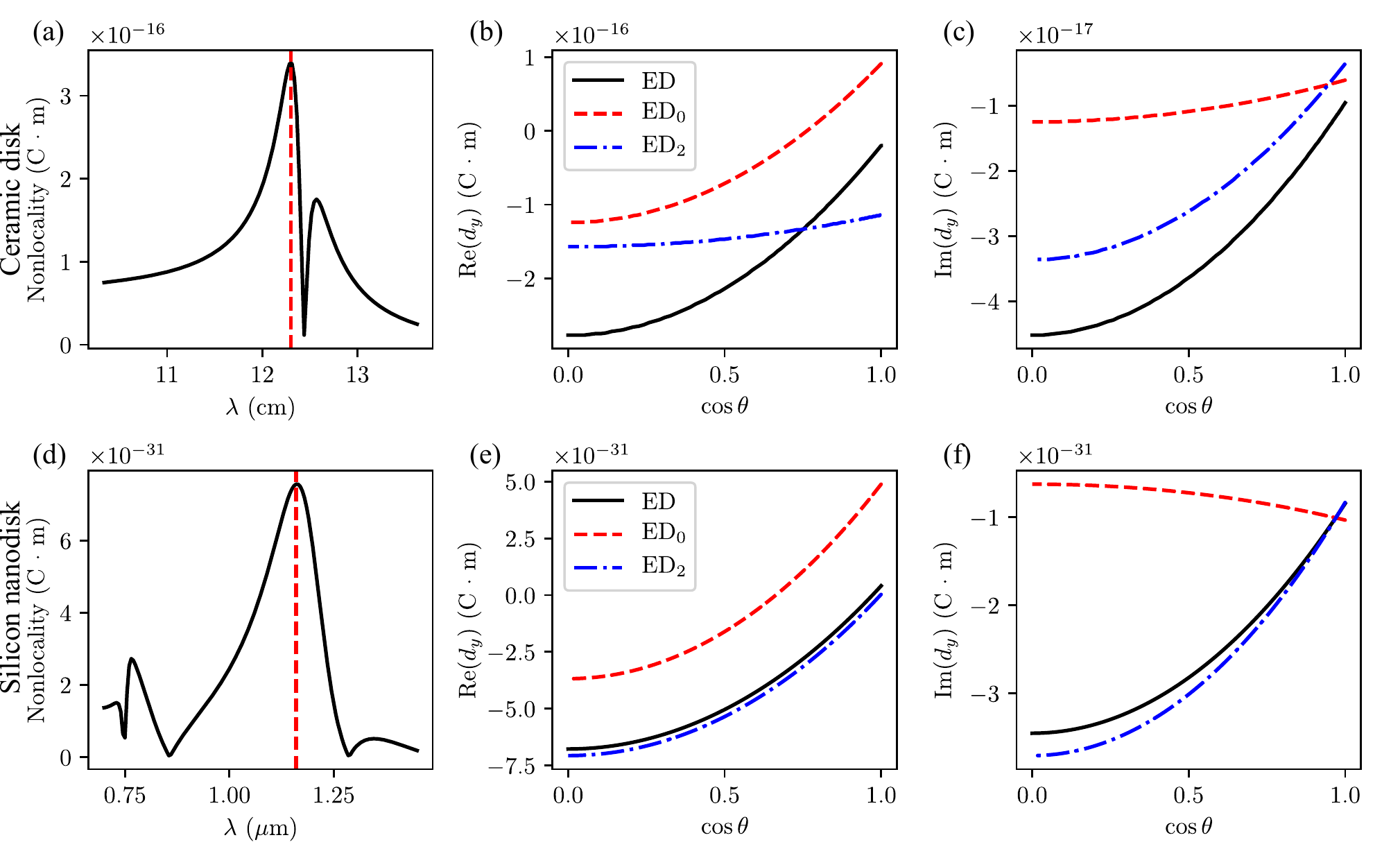}
	\caption{Dipole response of ceramic (a-c) and crystalline silicon (d-f) disks to the incident TE-polarized plane wave. (a,d) Spectral dependence of the nonlocal response quantified by the difference of the induced electric dipole moments $|d_y(\pi/2)-d_y(0)|$ for the two incidence angles $\theta=0$ and $\theta=\pi/2$. Red dashed line marks the frequency in the vicinity of electric anapole favouring the strongest nonlocal response. (b,c,e,f) Dependence of real (b,e) and imaginary (c,f) parts of induced dipole moment $d_y$ on $\cos\theta$ at wavelength (b,c) $\lambda=12.3$~cm; (e,f) $\lambda=1.18\,\mu$m. Black solid curve (ED) corresponds to the exact dipole moment calculated according to Refs.~\cite{Rockstuhl-OE-15,Rockstuhl-OC-18}. Red dashed line depicts the conventional dipole moment (ED$_0$) defined as ${\bf d}=\int\,{\bf P}\,dV$. Blue dot-dashed line shows dipole moment ED$_2$ calculated by Eq.~\eqref{Toroidal} incorporating the toroidal correction. All calculated dipole moments exhibit the dependence on the incidence angle.}
	\label{fig:Anapole}
	\end{figure}
	
\end{widetext}

First, we investigate the spectral dependence of this quantity [Fig.~\ref{fig:Anapole}(a,d)]. Our simulations suggest that thus defined nonlocality reaches its maximal value at the frequency close to the anapole minimum in the scattering spectra [Fig.~\ref{fig:spectrum}(a,c)]. We associate such behavior with strong suppression of local dipole response at the frequency of anapole when nonlocal effects become especially pronounced. 

Next we fix the wavelength of excitation to the value favouring the strongest nonlocal response and examine the dependence of the induced dipole moment $d_y$ on cosine of the incidence angle, $\cos\theta$. In agreement with Eq.~\eqref{dy}, this dependence is well-fitted by parabola [Fig.~\ref{fig:Anapole}(b,c,e,f)]. Furthermore, even zeroth-order approximation to the dipole moment ${\bf d}_0=\int {\bf P}\,dV$ (${\bf P}$ is electric polarization) exhibits the characteristic dependence on the incidence angle. At the same time, it strongly deviates from the full dipole moment defined according to Refs.~\cite{Rockstuhl-OE-15,Rockstuhl-OC-18}. Moreover, in the case of ceramic disk, the second-order approximation to the full dipole moment
\begin{equation}\label{Toroidal}
{\bf d}_2=\int\,{\bf P}\,dV-\frac{\omega^2}{10\,c^2}\,\int\,\left[2r^2\,{\bf P}-({\bf r}\cdot{\bf P})\,{\bf r}\right]\,dV
\end{equation}
also deviates from the exact result quite significantly despite the subwavelength size of the particle.

\begin{table}[h]
	\caption{\label{tab:table1} Polarizabilities and non-local parameters for ceramic disk at wavelength $\lambda=12.3$~cm}
	\begin{ruledtabular}
		\begin{tabular}{ccc}
			Parameter & Electric response & Magnetic response\\
			\colrule
			$\alpha_{\perp}$, cm$^3$ & $-73.73 + 57.26\,i$ & $-48.31 +117.74\,i$ \\
			$\alpha_{\parallel}$, cm$^3$ & $18.26 + 2.46\,i$ & $-15.11 + 1.69\,i$\\
			$\gamma_1$, cm$^5$ & $147.31 - 103.21\,i$ & $7.34 - 17.98\,i$\\
			$\gamma_2$, cm$^5$  & $-76.37 + 53.91\,i$ & $-8.24 + 20.13\,i$\\
			$\gamma_3$, cm$^5$  & $10.33 - 3.67\,i$ & $7.87 - 22.04\,i$\\
		\end{tabular}
	\end{ruledtabular}
\end{table}

The obtained results clearly indicate that the response of the disk is beyond the simplified model based on local polarizability tensors and the nonlocal corrections to the dipole moment provide a sizeable contribution to the total scattering cross-section at least in a certain frequency range.

Similarly to the dipole moment $d_y$, we extract the rest of the dipole moments excited in the disk by the incident TE or TM-polarized plane waves in geometry of Fig.~\ref{fig:disk}. Fitting the obtained dependence of the dipole moments on the incidence angle $\theta$, we retrieve all components of the particle polarizability tensor and associated nonlocal corrections as further discussed in Appendix~\ref{app:extraction}. For clarity, we focus on the case of a ceramic disk in which case the applicability of the dipole model is the most apparent. The extracted parameters are presented in Table~\ref{tab:table1}.

\begin{figure}[h!]
	\includegraphics{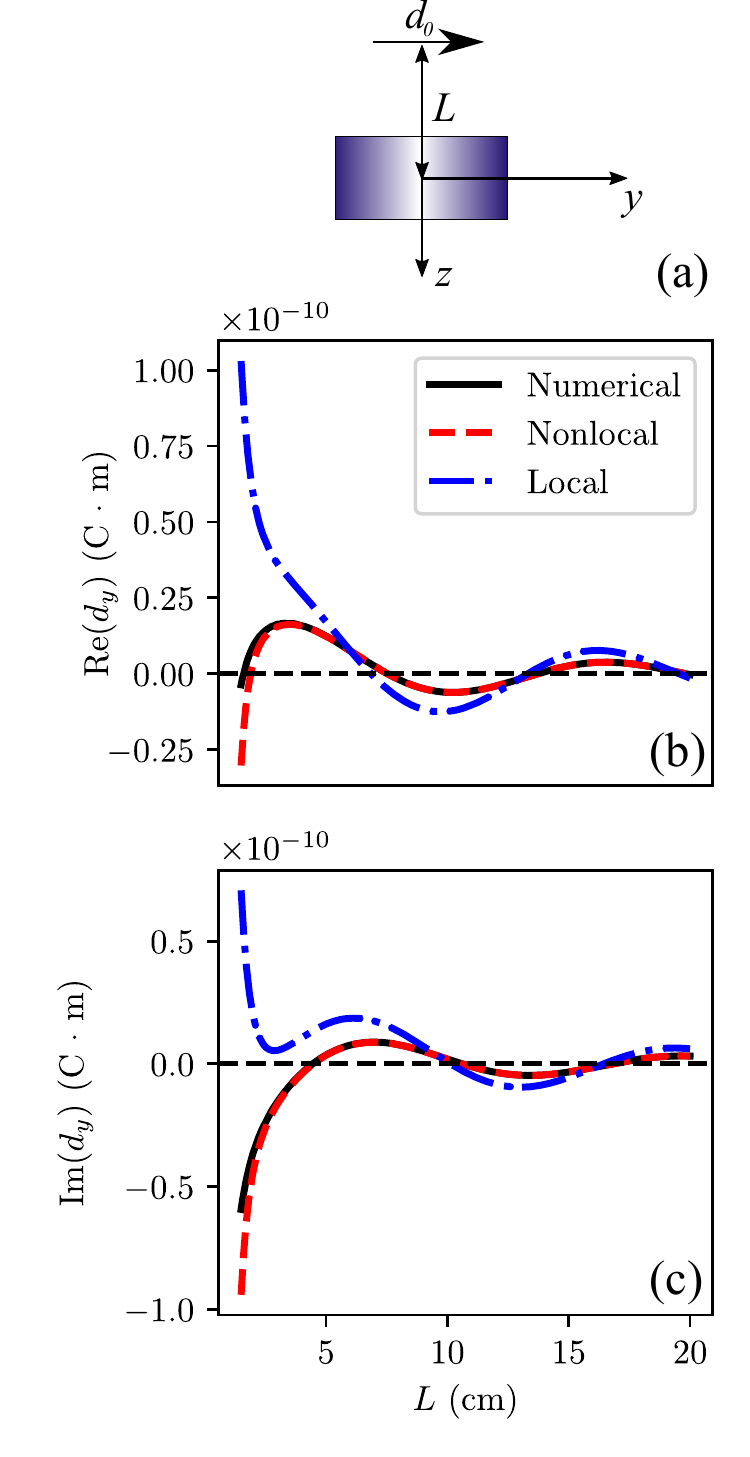}
	\caption{Excitation of the disk by a point electric dipole ${\bf d}_0$ placed at the symmetry axis of the disk. (a) Excitation geometry. (b,c) Real and imaginary parts of the dipole moment $d_y$ induced in the disk versus distance $L$ between point electric dipole ${\bf d}_0$ and the disk center. Black solid line shows the results of full-wave numerical simulations carried on in Comsol Multiphysics software package. Blue dot-dashed and red dashed  curves show the predictions of local and nonlocal models calculated using the extracted values of polarizability.}
	\label{fig:DipoleExcitation}
\end{figure}

Using the retrieved data, we check that the developed model captures the response of the disk not only to the propagating plane waves, but also to the evanescent near fields. To this end, we simulate the excitation of ceramic disk by a point electric dipole placed above the disk as illustrated in Fig.~\ref{fig:DipoleExcitation}(a) and oscillating at the same wavelength $\lambda=12.3$~cm matching to electric anapole. In this geometry, the field produced by the dipole at point ${\bf r}$ reads 
\begin{equation}
{\bf E}^{\rm{ext}}({\bf r})=\frac{1}{4\pi\,\eps_0}\,\hat{G}^{(\rm{ee})}(L\,\hat{z}+{\bf r})\,{\bf d}_0\:,
\end{equation}
where $\hat{G}^{\rm{(ee)}}$ is the dyadic Green's function describing the electric field produced by point electric dipole and $\hat{z}$ is a unit vector along $z$ axis.
%
%\begin{equation}
%\begin{aligned}
%\mathbf{E}^{\rm{ext}} = \dfrac{1}{4\pi \eps_0}\dfrac{e^{ikL}}{L^3} \left[ -1 + ikL + q^2 L^2 \right] d_0 \hat{y}\\
%+ \dfrac{1}{4\pi \eps_0}\dfrac{e^{ikL}}{L^3} \left[ 3 - 3ikL - q^2 L^2 \right] d_0 \hat{z}\:,
%\end{aligned}
%\end{equation}
%where $q = \om/c$, $\hat{y}$ and $\hat{z}$ are the unit vectors aligned along $y$ and $z$ axes, respectively. 
%
Using Eq.~\eqref{DipoleFull}, it is straightforward to evaluate the electric dipole moment induced in the disk:
\begin{equation}
d_y = \alpha_{\perp}\,\eps_0 E^{\rm{ext}}_y - \gamma_1\,\eps_0 \dfrac{\p^2 E^{\rm{ext}}_y}{\p z^2} - 2\gamma_2\,\eps_0\,\dfrac{\p^2 E^{\rm{ext}}_z}{\p y \p z}\:.
\end{equation}
The obtained results are shown by the red dashed line in Fig.~\ref{fig:DipoleExcitation}(b,c). At the same time, the prediction of the local model obtained by neglecting the nonlocal corrections is shown in Fig.~\ref{fig:DipoleExcitation}(b,c) by the blue dot-dashed curve. We observe a discrepancy between the two approaches evident at small distances $L<\lambda$, when the gradients of the field affecting the disk are especially large. To compare the two approaches, we extract the dipole moment of the disk $d_y$ directly from the full-wave numerical simulations as shown by the black solid curve in Fig.~\ref{fig:DipoleExcitation}(b,c). It is clearly seen that the nonlocal model perfectly agrees with full-wave simulations even at small distances.

\section{Describing the response of meta-crystals}\label{sec:Meta}

\begin{figure}[h!]
	\includegraphics[width=0.9\linewidth]{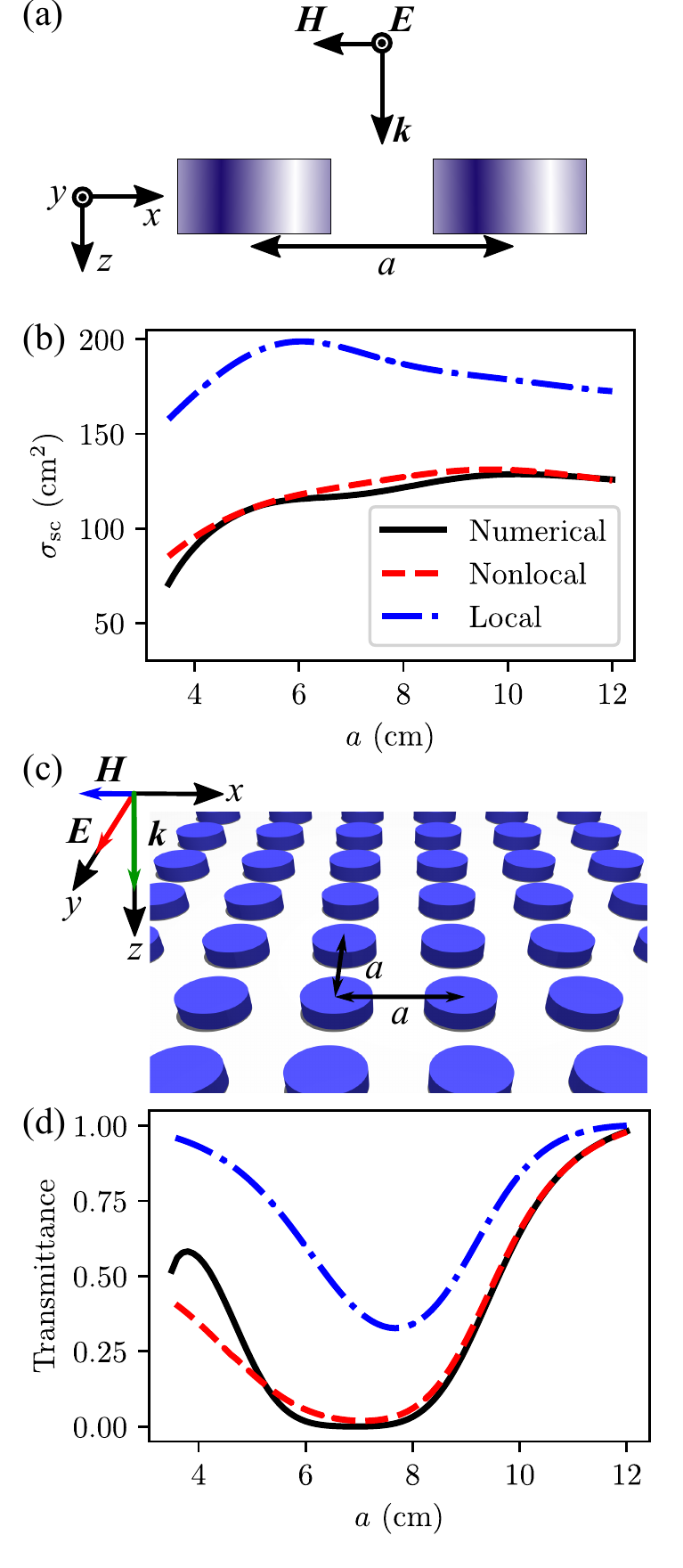}
	\caption{(a) Excitation of a pair of ceramic disks by the plane wave at normal incidence. (b) Scattering cross-section on a dimer versus distance $a$ between the disks. Blue dot-dashed and red dashed curves show the predictions of local and nonlocal models, respectively. Black solid line shows the results of full-wave numerical simulations. (c) Excitation of a metasurface based on the square lattice of ceramic disks with the period $a$ by the plane wave at normal incidence. (d) The dependence of metasurface transmittance on the period $a$. Blue dot-dashed and red dashed curves show the predictions of local and nonlocal models, respectively. Black solid line shows the results of full-wave numerical simulations.}
	\label{fig:DiskInteraction}
\end{figure}

Since the developed model of the disk nonlocal response includes only few extra parameters, it can be readily applied to describe the clusters composed of such disks as well as metasurfaces. First we examine a relatively simple case when the incident plane wave with the wavelength $\lambda=12.3$~cm scatters on a dimer composed of the two identical disks [Fig.~\ref{fig:DiskInteraction}(a)]. To describe such system, we write down the self-consistent equations for the dipole moments of the disks taking into account their mutual interaction. Once the dipole moments of the disks are found, the scattered radiation and the scattering cross-section can be readily evaluated. The predictions of local and nonlocal models are depicted in Fig.~\ref{fig:DiskInteraction}(b) showing a significant discrepancy. Comparing these results with full-wave numerical simulations, we observe that the nonlocal model fits numerical results much better. Nevertheless, slight discrepancies are present.

The reason for such discrepancies is related to the near-field interaction of the disks. Even though higher-order multipoles are off-resonant at wavelength of interest, gradients of the near fields do excite such multipoles. As a consequence, higher-order multipoles provide a contribution to the scattering cross-section which is especially pronounced for small distances $a$.

Having tested our approach on a simple problem of a dimer, we now switch to more interesting scenario of a metasurface based on the square lattice of ceramic disks with the period $a$ [Fig.~\ref{fig:DiskInteraction}(c)]. For simplicity, we focus on the geometry of normal incidence, in which case only $d_y$ and $m_x$ components of the dipole moments are excited. The developed model suggests that the dipole moments are defined by the local fields acting on the disks:
\begin{gather}
d_y=\alpha_{\bot}\,\eps_0\,E_y-\gamma_1\,\eps_0\,\ds{E_y}{z}-2\gamma_2\,\eps_0\,\frac{\partial^2 E_z}{\partial y\,\partial z}\:,\label{eq:dymeta}\\
m_x=\alpha_{\bot}^{\rm{(m)}}\,H_x-\gamma_1^{{\rm (m)}}\,\ds{H_x}{z}-2\gamma_2^{{\rm (m)}}\,\frac{\partial^2 H_z}{\partial x\,\partial z}\:.\label{eq:mxmeta}
\end{gather}
The local fields are in turn presented as a superposition of the incident fields $E_{\rm{in}}$, $H_{\rm{in}}$ and the fields scattered by the rest of the particles in the array:
\begin{gather}
E_y=E_{{\rm in}}\,e^{iqz}+\frac{1}{4\pi\,\eps_0}\,G_{kyy}^{{\rm (ee)}}\,d_y+\frac{Z_0}{4\pi}\,G_{kyx}^{{\rm (em)}}\,m_x\:,\label{eq:Ey}\\
H_x=-H_{{\rm in}}\,e^{iqz}+\frac{c}{4\pi}\,G_{kxy}^{{\rm (me)}}\,d_y+\frac{1}{4\pi}\,G_{kxx}^{{\rm (mm)}}\,m_x\:,\\
E_z=\frac{1}{4\pi\,\eps_0}\,G_{kzy}^{\rm{(ee)}}\,d_y+\frac{Z_0}{4\pi}\,G_{kzx}^{\rm{(em)}}\,m_x\:,\\
H_z=\frac{c}{4\pi}\,G_{kzy}^{\rm{(me)}}\,d_y+\frac{1}{4\pi}\,G_{kzx}^{\rm{(mm)}}\,m_x\:.\label{eq:Hz}
\end{gather}
Here, $Z_0=\sqrt{\mu_0/\eps_0}$ is the free space impedance and $G_{kij}^{\rm{(\alpha \beta)}}$ are lattice sums which describe the field produced by all particles of the array except of the given one:
\begin{equation}
G_{kij}^{\rm{(\alpha \beta)}}=\sum\limits_{(m,n)\not=(0,0)}\,G_{ij}^{(\alpha \beta)}(-{\bf r}_{mn})\,e^{i{\bf k}\cdot{\bf r}_{mn}}\:,
\end{equation}
where ${\bf k}$ denotes an in-plane wave vector, $G$ is the dyadic Green's function, $(i,j)$ are its Cartesian components, and the upper indices $\alpha$ and $\beta$ indicate the type of the dyadic Green's function, for instance, ${\rm (ee)}$ and ${\rm (em)}$ stand for electric field produced by electric and magnetic dipoles, respectively.

Combining Eqs.~\eqref{eq:dymeta}-\eqref{eq:Hz}, we derive the expressions for the dipole moments. Taking into account that the lattice sums containing $z$-derivatives of the odd order vanish and $\hat{G}^{\rm{(mm)}}=\hat{G}^{\rm{(ee)}}$, we calculate the dipole moments of the particles:
\begin{gather}
d_y=(\alpha_{\bot}+\gamma_1\,q^2)\,\eps_0\,E_{\rm{in}}\times\notag\\
\left[1-\frac{\alpha_{\bot}}{4\pi}\,G_{kyy}+\frac{\gamma_1}{4\pi}\,\ds{G_{kyy}}{z}+\frac{\gamma_2}{2\pi}\,\frac{\partial^2 G_{kzy}}{\partial y\,\partial z}\right]^{-1}\:,\\
m_x=-(\alpha_{\bot}^{\rm{(m)}}+\gamma_1^{\rm{(m)}}\,q^2)\,H_{{\rm in}}\times\notag\\
\left[1-\frac{\alpha_{\bot}^{\rm{m}}}{4\pi}\,G_{kxx}+\frac{\gamma_1^{\rm{m}}}{4\pi}\,\ds{G_{kxx}}{z}+\frac{\gamma_2^{\rm{m}}}{2\pi}\,\frac{\partial^2 G_{kzx}}{\partial x\,\partial z}\right]^{-1}\:,
\end{gather}
where the upper ${\rm (ee)}$ index of the Green's functions is suppressed throughout for brevity. The transmitted field is obtained by summing the far fields produced by all particles comprising the metasurface~\cite{Belov2006}:
\begin{equation}
E_{{\rm t}}=E_{\rm{in}}+\frac{i\,q}{2\eps_0\,a^2}\,(d_y-m_x/c)\:.
\end{equation}
The most involved part of this calculation is the evaluation of the lattice sums governing the interaction of the particles within the metasurface, and this part is discussed in Appendix~\ref{app:Sums}.

Calculated results for metasurface transmittance are presented in Fig.~\ref{fig:DiskInteraction}(d). The nonlocal model perfectly matches the results of full-wave numerical simulations once the period of the metasurface $a$ is larger than $(8\div 9)$~cm, i.e. exceeds the diameter of the disk approximately three times. For shorter distances, the agreement becomes worse, which is related to the intrinsic limitations of the discrete dipole model and agrees with the other studies~\cite{Chebykin}. Similarly to the case of dimer, this discrepancy is associated with the excitation of higher-order multipoles in the disk by the gradients of the near fields.

\section{Discussion and outlook}\label{sec:Concl}

In summary, we have investigated dipole response of Mie-resonant non-spherical particles in the region of crossover from $R/\lambda\ll 1$ to $R/\lambda\sim 1$, which is the case for experimentally relevant situations. As we have proved for the case of dielectric disks, induced dipole moments are determined not only by the fields in the particle center but also by the second-order spatial derivatives of the fields, which gives rise to nonlocality of the particle dipole response. We have also demonstrated that the predicted nonlocal effects are especially pronounced in the vicinity of anapole minimum in the scattering spectrum reaching up to 50\% of local response thereby largely governing light scattering in this frequency range.

While the developed model includes only few extra parameters describing the nonlocal effects, it captures the response of the disk not only to the propagating plane waves but also to the evanscent fields. Moreover, our approach is applicable also to the metamaterials and metasurfaces composed of Mie-resonant scatterers, providing an improved accuracy in comparison with the standard discrete dipole approximation.

As we prove, our results are universally valid across the entire electromagnetic spectrum, being applicable not only in the microwave domain but also at infrared and visible frequencies. The approach developed here can be directly generalized to the cases of less symmetric particles or larger scatterers when higher-order spatial derivatives of the field should be taken into account. Moreover, our analysis can be also applied to the case of higher-order multipole moments, for instance, electric and magnetic quadrupoles.

Nonlocality of the dipole response provides an interesting perspective on spatial dispersion effects in metamaterials, which are normally described via the expansion of permittivity tensor in powers of the wave vector~\cite{Agranovich,Silveirinha2007,Mnasri2018}. In the previous microscopic descriptions, nonlocality in metamaterials has been linked to the interaction of resonant scatterers with each other~\cite{Silv-L-2007,Gorlach-14,Chebykin}. Now it is apparent that this picture should be supplemented by the nonlocal response of the individual particle.

Finally, we believe that our findings provide valuable insights into meta-optics and all-dielectric nanophotonics by highlighting truly nonlocal behavior of their basic building blocks~--- Mie-resonant nanoparticles.

{\it Note added in proof.} Recently, we became aware of a complementary investigation of retardation effects in the dipole response of nanostructures~\cite{Arbouet}.

\begin{acknowledgments}
We acknowledge Pavel Belov and Kseniia Baryshnikova for valuable discussions. Theoretical models were supported by the Russian Science Foundation (Grant No.~20-72-10065), numerical simulations were supported by the Russian Foundation for Basic Research (Grant No.~18-32-20065). M.A.G. acknowledges partial support by the Foundation for the Advancement of Theoretical Physics and Mathematics ``Basis''. D.A.S. acknowledges support by the Australian Research Council (grant DE190100430).
\end{acknowledgments}

\appendix

\section{Symmetry analysis of the disk dipole response}\label{app:symmetry}

In this Appendix, we  calculate the number of independent components of the tensors $\alpha_{ij}$ and $\gamma_{ijlm}$ that enter Eq.~\eqref{eq1} applying group-theoretical arguments~\cite{Dresselhaus}. We assume $D_{\infty h}$ symmetry group of the particle and take into account symmetry of the tensors with respect to permutation of $i,j$ and $l,m$ indices.

First, we notice that the vectors ${\bf d}$, ${\bf E}$ and ${\bf k}$ transform according to $E_{1u}(\Pi_u)$ representation of $D_{\infty h}$ symmetry group. To simplify our analysis, we consider finite $D_{Nh}$ group setting $N\rightarrow\infty$ at later steps. This group includes rotations around vertical $z$ axis by angles $\varphi_n=2\pi\,n/N$, where $n=0\dots (N-1)$, further denoted as $C_{\varphi}$;  rotations by $\pi$ around $N$ horizontal symmetry axes of polygon, $C_2'$; all previous symmetry transformations followed by spatial inversion, $i\,C_{\varphi}$ and $i\,C_2'$. Note that identity element is contained in $C_{\varphi}$ with $\varphi=0$. The characters of all these transformations for $E_{1u}(\Pi_u)$ representation are provided in the third line of Table~\ref{tab:table2}.

\begin{table}[b]
	\caption{\label{tab:table2} Character table for $E_{1u}(\Pi_u)$ representation of $D_{Nh}$ symmetry group in 3D case}
	\begin{tabular}{|c|cccc|}
%		\begin{tabular}
            \colrule
			$R$ & $C_{\varphi}$ & $C_2'$ & $i\,C_{\varphi}$ & $i\,C_2'$\\
			\colrule
			$N_k$ & $N$ & $N$ & $N$ & $N$\\
			$\chi(R)$ & $1+2 \cos\varphi$ & -1 & $-1-2 \cos\varphi$ & 1\\
			$\chi(R^2)$ & $1+2 \cos2\varphi$ & 3 & $1+2 \cos2\varphi$ & 3\\
			$\left<\overline{\chi}(R)\right>_{\varphi}$ & 2 & 2 & 2 & 2\\
			$\left<\tilde{\chi}(R)\right>_{\varphi}$ & 8 & 4 & 8 & 4\\
			\colrule
%		\end{tabular}
	\end{tabular}
\end{table}

Without additional constraints, representations of the tensors $\alpha_{ij}$ and $\gamma_{ijlm}$ correspond to the tensor products $\mathcal{D}\otimes \mathcal{D}$ and $\mathcal{D}\otimes \mathcal{D}\otimes \mathcal{D}\otimes \mathcal{D}$, respectively, where $\mathcal{D}$ matrices realize $E_{1u}(\Pi_u)$ representation of $D_{Nh}$ symmetry group. However, constructed matrix representations should also be symmetric with respect to the interchange of indices $i,j$ and $l,m$. Hence, we should construct symmetrized matrices of representation
\begin{gather}
\bar{\mathcal{D}}_{ij,i'j'}(R)=\frac{1}{2}\,\left[\mathcal{D}_{ii'}\,\mathcal{D}_{jj'}+\mathcal{D}_{ij'}\,D_{ji'}\right]\:,\\
\bar{\mathcal{D}}_{ijlm,i'j'l'm'}(R)\notag\\
=\frac{1}{4}\,\left[\mathcal{D}_{ii'}\,\mathcal{D}_{jj'}\,\mathcal{D}_{ll'}\,\mathcal{D}_{mm'}
+\mathcal{D}_{ij'}\,\mathcal{D}_{ji'}\,\mathcal{D}_{ll'}\,\mathcal{D}_{mm'}\right.\\
\left.+\mathcal{D}_{ii'}\,\mathcal{D}_{jj'}\,\mathcal{D}_{lm'}\,\mathcal{D}_{ml'}+\mathcal{D}_{ij'}\,\mathcal{D}_{ji'}\,\mathcal{D}_{lm'}\,\mathcal{D}_{ml'}\right]\:.\notag
\end{gather}
The respective characters of these symmetrized representations are calculated as:
\begin{gather}
\bar{\chi}(R)=\frac{1}{2}\,\left[\chi^2(R)+\chi(R^2)\right]\:,\\
\tilde{\chi}(R)=\frac{1}{4}\,\left[\chi^4(R)+2\,\chi^2(R)\,\chi(R^2)+\chi^2(R^2)\right]\:.
\end{gather}
Clearly, $\bar{\chi}$ and $\tilde{\chi}$ calculated for $C_{\varphi}$ and $i\,C_{\varphi}$ depend on $\varphi$, but in our analysis we are interested in the values of these characters averaged over $\varphi$; these values are provided in the fifth and sixth lines of Table~\ref{tab:table2}.

Finally, we can determine the number of independent components of the tensors under consideration by calculating the number of times that unity representation enters the constructed symmetrized representations:
\begin{equation}
\nu(\alpha)=\frac{1}{4\,N}\,\sum\,N_k\,\bar{\chi}(C_k)=2\:,
\end{equation}

\begin{equation}
\nu(\gamma)=\frac{1}{4\,N}\,\sum\,N_k\,\tilde{\chi}(C_k)=6\:.
\end{equation}

Hence, $\alpha_{ij}$ and $\gamma_{ijlm}$ tensors contain two and six independent components, respectively, as indicated in the article main text.

\section{Retrieving the nonlocal corrections to the particle polarizability tensor}\label{app:extraction}

\begin{widetext}
	
\begin{figure}[t]
	\includegraphics{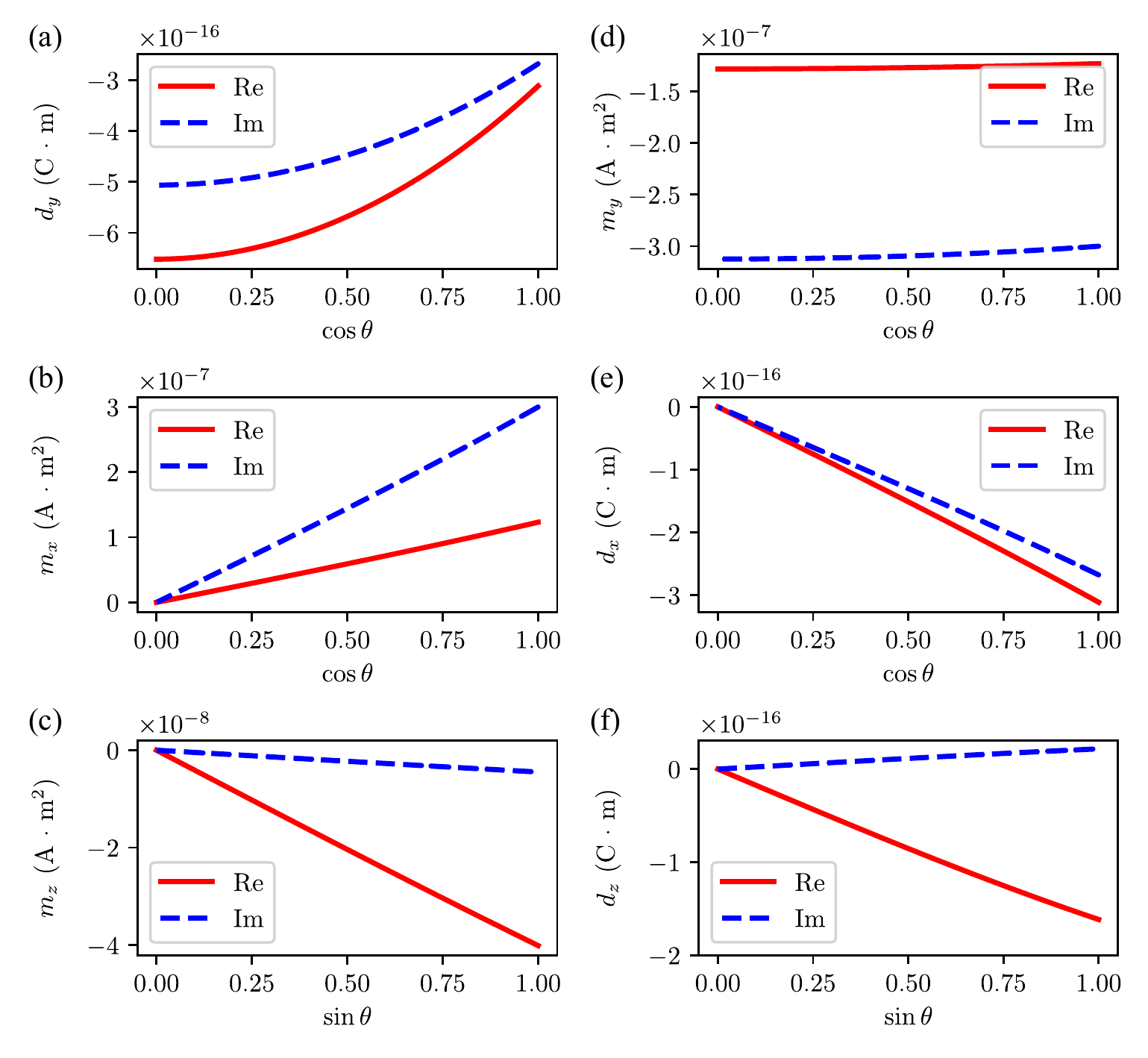}
	\caption{Real and imaginary parts of dipole moments induced by the incident TE- (a,b,c) and TM- (d,e,f) polarized plane wave as a function of the incidence angle $\theta$ in geometry of Fig.~\ref{fig:disk}. (a) Electric dipole, $y$-component. (b) Magnetic dipole, $x$-component. (c) Magnetic dipole, $z$-component. (d) Magnetic dipole, $y$-component. (e) Electric dipole, $x$-component. (f) Electric dipole, $z$-component.}
	\label{fig:dipoles_tetm}
\end{figure}

\end{widetext}

Developing the retrieval procedure for the nonlocal dipole response of the disk, we focus on the case of ceramic disk, since in this case the dipole resonances are well-isolated spectrally. We fix excitation wavelength to $\lambda=12.3$~cm which corresponds to the excitation frequency $f=2.44$~GHz and vary the incidence angle $\theta$ studying the geometry Fig.~\ref{fig:disk}. Using multipole decomposition technique~\cite{Rockstuhl-OE-15,Rockstuhl-OC-18}, we evaluate complex amplitudes of the disk dipole moments for both polarizations of excitation as a function of $\theta$. Setting the phase of the incident wave to zero in the disk center, we plot real and imaginary parts of the retrieved dipole moments for TE and TM-polarized excitations in Fig.~\ref{fig:dipoles_tetm}.

Dipole moment $d_y$ induced by TE-polarized plane wave features strong dependence on the incidence angle shown in Fig.~\ref{fig:dipoles_tetm}(a). Since the disk is non-bianisotropic, this dipole moment arises purely due to the electric field ${\bf E}$. However, $y$ projection of the electric field does not depend on the incidence angle $\theta$ and therefore the results of simulations clearly show that the response of the disk is indeed beyond the simplified model based on local polarizability tensors. Moreover, the dipole moment changes by more than 50\% when the incidence angle is varied from 0 to $\pi/2$ which hints towards strong nonlocal response at the chosen frequency.

Somewhat similar situation is observed for TM-polarized excitation, when induced magnetic moment $m_y$ also depends on the incidence angle [Fig.~\ref{fig:dipoles_tetm}(d)], serving as a fingerprint of ``magnetic'' nonlocalities. However, they appear to be significantly smaller than their electric counterparts which is explained by the choice of the frequency. For instance, tuning the excitation frequency to the so-called magnetic anapole when magnetic dipole radiation is cancelled, one can expect a substantial enhancement of ``magnetic'' nonlocalities.

The rest of dipole moments [Fig.~\ref{fig:dipoles_tetm}(b,c,e,f)] feature almost linear dependence on $\cos\theta$ or $\sin\theta$ with relatively small deviations in agreement with Eqs.~\eqref{mx}-\eqref{dz}. To provide more comprehensive picture, we fit the simulation data by Eqs.~\eqref{dy}-\eqref{dz}, extracting the relevant parameters: 5 for electric and 5 for magnetic dipole response of the disk. Calculated results are presented in Table~\ref{tab:table1}.

Chosen frequency close to the resonance for in-plane magnetic dipoles gives rise to a quite large imaginary part of magnetic polarizability $\alpha_{\perp}^{(\rm{m})}$. Real parts of polarizabilities $\alpha_{\parallel}^{(\rm{m})}$, $\alpha_{\perp}^{(\rm{m})}$ and $\alpha_{\perp}^{(\rm{m})}$ appear to be negative, since the considered frequency $2.44$~GHz is higher than the resonance frequencies for $z$-oriented magnetic dipole or in-plane dipoles. At the same time, the frequency of excitation is below the resonance frequency for $z$-oriented electric dipole, which ensures that the real part of $\alpha_{\parallel}$ is positive.

\section{Lattice sums for the metasurface}\label{app:Sums}

In this Appendix, we discuss the calculation of the two-dimensional lattice sums necessary to evaluate the transmission and reflection coefficients for a metasurface at normal incidence within generalized discrete dipole approach. We employ the technique~\cite{Belov2005} based on Poisson summation formula, extending it to calculate the sum of the spatial derivatives of the Green's function.

The sum $G_{kxx}$ has been calculated in Ref.~\cite{Belov2005} and for $e^{-i\omega\,t}$ time convention is given by the formula:

\begin{widetext}
\begin{gather}
G_{kxx}=4\pi\,(A_1^*+A_{2a}^*+A_{2b}^*)\:,\label{eq:Gkxx}\\
A_1=\frac{1}{\pi\,a^3}\,\sum\limits_{m=1}^{\infty} \frac{(2i\,q\,a+3)\,m+2}{m^3\,(m+1)\,(m+2)}\,e^{-iq\,m\,a}
\cos(k_x\,m\,a)\notag\\
+\frac{1}{4\pi\,a^3}\,\left[-(iq\,a+1)\,\left(t_{+}^2\,\ln t^{+}+t_{-}^2\,\ln t^{-}+2\,e^{iqa}\,\cos k_x\,a\right)-2iq\,a\,\left(t_{+}\,\ln t^{+}+t_{-}\,\ln t^{-}\right)+(7i\,q\,a+3)\right]\:,\\
A_{2a}=-\sum\limits_{\text{Re}\,p_m\not=0}\,\frac{p_m^2}{\pi\,a}\,\sum\limits_{n=1}^{\infty}\,K_0(p_m\,a\,n)\,\cos(k_y\,a\,n)\:,\\
A_{2b}=-\sum\limits_{\text{Re}\,p_m=0}\,\frac{p_m^2}{2\,a^2}\,\left\lbrace\frac{1}{\sqrt{k_y^2+p_m^2}}+\zeta(3)\,\frac{(2k_y^2-p_m^2)\,a^3}{8\,\pi^3}+\frac{a}{\pi}\,\left[\ln\left(\frac{a|p_m|}{4\pi}\right)+\frac{ia}{2}\right]\right\rbrace\notag\\
-\sum\limits_{\text{Re}\,p_m=0}\,\frac{p_m^2}{2\,a^2}\,\sum\limits_{n=1}^{\infty}\,\left[\frac{1}{\sqrt{(k_y+2\pi\,n/a)^2+p_m^2}}+\frac{1}{\sqrt{(k_y-2\pi\,n/a)^2+p_m^2}}-\frac{a}{\pi\,n}-\frac{(2k_y^2-p_m^2)\,a^3}{8 \pi^3\,n^3}\right]\:.\label{eq:A2b}
\end{gather}
Here $p_m=\sqrt{(k_x+2\pi m/a)^2-q^2}$. If $\text{Re}\,p_m\not=0$, we choose $+$ sign in front of the square root, otherwise we calculate $p_m$ as $i\,\sqrt{q^2-(k_x+2\pi m/a)^2}$. Square roots comprising Eq.~\eqref{eq:A2b} are understood in the same way. $t^{\pm}=1-e^{-i(q\pm k_x)\,a}$, $t_{\pm}=1-e^{i(q\pm k_x)\,a}$. In the case of normal incidence $k_x=k_y=0$. The series $A_1$ includes the fields of dipoles oriented along $x$ axis and has a power-law convergence. The series $A_{2a}$ ($A_{2b}$) are associated with the evanescent (propagating) fields produced by the rest of the particles and feature exponential (power-law) convergence. Overall, Eqs.~\eqref{eq:Gkxx}-\eqref{eq:A2b} are quite suitable for rapid numerical calculations.

To calculate the derivative $\partial^2 G_{kxx}/\partial z^2$ we introduce the nonzero $z$ coordinate of the observation point and take the second derivative of the lattice sum $G_{kxx}$ with respect to $z$. The result reads:
\begin{gather}
\ds{G_{kxx}}{z}=4\pi\,(B_1^*+B_{2a}^*+B_{2b}^*)\:,\label{eq:Gkxxzz}\\
B_1=-\frac{q^2}{4\pi\,a^3}\,\left(t_{+}^2\,\ln t^{+}-t_{+}+t_{-}^2\,\ln t^{-}-t_{-}-1\right)\notag\\
+\frac{1}{\pi\,a^5}\,\sum\limits_{m=1}^{\infty}\,\cos\left(k_x\,m\,a\right)\,e^{-iq\,m\,a}\,\left(-\frac{3}{m^5}-\frac{3iq\,a}{m^4}+\frac{q^2\,a^2\,(3m+2)}{m^3\,(m+1)\,(m+2)}\right)\:,\\
B_{2a}=\sum\limits_{\text{Re}\, p_m\not=0}\,\sum\limits_{n=1}^{\infty}\,\frac{p_m^3}{\pi\,a^2}\,\frac{K_1(p_m\,a\,n)}{n}\,\cos(k_y\,a\,n)\:,\\
B_{2b}=\sum\limits_{\text{Re}\, p_m=0}\,\frac{p_m^3}{\pi\,a^2}\,\sum\limits_{n=1}^{\infty}\,\frac{\cos(k_y\,a\,n)}{n}\,\left[K_1(p_m\,a\,n)-\sqrt{\frac{\pi}{2\,p_m\,a\,n}}\,e^{-p_m\,a\,n}\,\left(1+\frac{3}{8\,p_m\,a\,n}-\frac{15}{128\,(p_m\,a\,n)^2}\right)\right]\notag
\end{gather}

\begin{gather}
+\sum\limits_{\text{Re}\, p_m=0}\,\frac{p_m^4}{\sqrt{8\pi}\,a}\,\left[\frac{1}{(p_m\,a)^{3/2}}\,\left(\pol{3/2,r_{+}}+\pol{3/2,r_{-}}\right)\right.\\
\left.+\frac{3}{8\,(p_m\,a)^{5/2}}\,\left(\pol{5/2,r_{+}}+\pol{5/2,r_{-}}\right)-\frac{15}{128\,(p_m\,a)^{7/2}}\,\left(\pol{7/2,r_{+}}+\pol{7/2,r_{-}}\right)\right]\:,\notag\label{eq:B2b}
\end{gather}
where the same designations are used, $r_{\pm}=\exp(-p_m\,a\pm ik_y\,a)$ and $\text{polylog}$ function is defined as:
\begin{equation*}
\pol{a,z}=\sum\limits_{n=1}^{\infty}\,\frac{z^n}{n^a}\:.
\end{equation*}
In a similar way we calculate another derivative of the lattice sum:
\begin{gather}
\frac{\partial^2\,G_{kyz}}{\partial y\,\partial z}=C_1+C_{2a}+C_{2b}\:,\\
C_1=-\frac{iq^3}{a^2}\,\left(-\tau_{+}\,\ln \tau^{+}-\tau_{-}\,\ln \tau^{-}+2\right)+\left(\frac{iq^3}{2\,a^2}-\frac{5q^2}{2\,a^3}\right)\,\left(\tau_{+}^2\,\ln \tau^{+}-\tau_{+}+\tau_{-}^2\,\ln \tau^{-}-\tau_{-}-1\right)\notag\\
-\frac{2iq^3}{a^2}\,\sum\limits_{n=1}^{\infty}\,\frac{\rho_{+}^n+\rho_{-}^n}{n^2\,(n+1)\,(n+2)}+\frac{5\,q^2}{a^3}\,\sum\limits_{n=1}^{\infty}\,\frac{(\rho_{+}^n+\rho_{-}^n)\,(3n+2)}{n^3\,(n+1)\,(n+2)}+\sum\limits_{n=1}^{\infty}\,\frac{12\,(iq\,n\,a-1)}{n^5\,a^5}\,(\rho_{+}^n+\rho_{-}^n)\:,\\
C_{2a}=\frac{4}{a}\,\sum\limits_{\text{Re}\,\kappa_n\not=0}\,\sum\limits_{m=1}^{\infty}\,\frac{\cos (k_x\,m\,a)}{m\,a}\,\kappa_n\,(k_y^{(n)})^2\,K_1(m\,a\,\kappa_n)\:,\\
C_{2b}=\sum\limits_{\text{Re}\,\kappa_n=0}\,\frac{\sqrt{2\pi}\kappa_n^2\,(k_y^{(n)})^2}{a}\,\left[\frac{1}{(\kappa_n\,a)^{3/2}}\,\left(\pol{3/2,s_{+}}+\pol{3/2,s_{-}}\right)\right.\notag\\
\left.+\frac{3}{8\,(\kappa_n\,a)^{5/2}}\,\left(\pol{5/2,s_{+}}+\pol{5/2,s_{-}}\right)-\frac{15}{128 (\kappa_n\,a)^{7/2}}\,\left(\pol{7/2,s_{+}}+\pol{7/2,s_{-}}\right)\right]\notag\\
+\frac{4}{a}\sum\limits_{\text{Re}\,\kappa_n=0}\,\sum\limits_{m=1}^{\infty}\,\kappa_n\,(k_y^{(n)})^2\,\frac{\cos k_x\,m\,a}{m\,a}\,\left[K_1(\kappa_n\,m\,a)-\sqrt{\frac{\pi}{2\,\kappa_n\,m\,a}}\,e^{-\kappa_n\,m\,a}\,\left(1+\frac{3}{8\,\kappa_n\,m\,a}-\frac{15}{128\,(\kappa_n\,m\,a)^2}\right)\right]\:,
\end{gather} 
where $\rho_{\pm}=e^{i(q\pm k_y)\,a}$, $\tau^{\pm}=1-\rho_{\pm}$, $\tau_{\pm}=1-e^{-i(q\pm k_y)\,a}$, $k_y^{(n)}=2\pi\,n/a+k_y$, $\kappa_n=\sqrt{(k_y^{(n)})^2-q^2}$. If $\text{Re}\,\kappa_n\not=0$, we choose $+$ sign in front of the square root. Otherwise, if $\text{Re}\,\kappa_n=0$, we calculate $\kappa_n$ as $-i\,\sqrt{q^2-(k_y^{(n)})^2}$. $s_{\pm}=e^{-\kappa_n\,a\pm ik_x\,a}$.

Note that in the geometry of normal incidence ($k_x=k_y=0$) $G_{kyy}=G_{kxx}$, $\partial^2\,G_{kyy}/\partial z^2=\partial^2\,G_{kxx}/\partial z^2$, $\partial^2\,G_{kyz}/(\partial y\,\partial z)=\partial^2\,G_{kxz}/(\partial x\,\partial z)$ due to $D_4$ symmetry of a metasurface. Hence, only three independent lattice sums have to be evaluated.

\end{widetext}

\bibliography{NonlocalityLib}

\begin{thebibliography}{36}%
\makeatletter
\providecommand \@ifxundefined [1]{%
 \@ifx{#1\undefined}
}%
\providecommand \@ifnum [1]{%
 \ifnum #1\expandafter \@firstoftwo
 \else \expandafter \@secondoftwo
 \fi
}%
\providecommand \@ifx [1]{%
 \ifx #1\expandafter \@firstoftwo
 \else \expandafter \@secondoftwo
 \fi
}%
\providecommand \natexlab [1]{#1}%
\providecommand \enquote  [1]{``#1''}%
\providecommand \bibnamefont  [1]{#1}%
\providecommand \bibfnamefont [1]{#1}%
\providecommand \citenamefont [1]{#1}%
\providecommand \href@noop [0]{\@secondoftwo}%
\providecommand \href [0]{\begingroup \@sanitize@url \@href}%
\providecommand \@href[1]{\@@startlink{#1}\@@href}%
\providecommand \@@href[1]{\endgroup#1\@@endlink}%
\providecommand \@sanitize@url [0]{\catcode `\\12\catcode `\$12\catcode
  `\&12\catcode `\#12\catcode `\^12\catcode `\_12\catcode `\%12\relax}%
\providecommand \@@startlink[1]{}%
\providecommand \@@endlink[0]{}%
\providecommand \url  [0]{\begingroup\@sanitize@url \@url }%
\providecommand \@url [1]{\endgroup\@href {#1}{\urlprefix }}%
\providecommand \urlprefix  [0]{URL }%
\providecommand \Eprint [0]{\href }%
\providecommand \doibase [0]{https://doi.org/}%
\providecommand \selectlanguage [0]{\@gobble}%
\providecommand \bibinfo  [0]{\@secondoftwo}%
\providecommand \bibfield  [0]{\@secondoftwo}%
\providecommand \translation [1]{[#1]}%
\providecommand \BibitemOpen [0]{}%
\providecommand \bibitemStop [0]{}%
\providecommand \bibitemNoStop [0]{.\EOS\space}%
\providecommand \EOS [0]{\spacefactor3000\relax}%
\providecommand \BibitemShut  [1]{\csname bibitem#1\endcsname}%
\let\auto@bib@innerbib\@empty
%</preamble>
\bibitem [{\citenamefont {Kuznetsov}\ \emph {et~al.}(2016)\citenamefont
  {Kuznetsov}, \citenamefont {Miroshnichenko}, \citenamefont {Brongersma},
  \citenamefont {Kivshar},\ and\ \citenamefont
  {Luk'yanchuk}}]{Kuznetsov-Science}%
  \BibitemOpen
  \bibfield  {author} {\bibinfo {author} {\bibfnamefont {A.~I.}\ \bibnamefont
  {Kuznetsov}}, \bibinfo {author} {\bibfnamefont {A.~E.}\ \bibnamefont
  {Miroshnichenko}}, \bibinfo {author} {\bibfnamefont {M.~L.}\ \bibnamefont
  {Brongersma}}, \bibinfo {author} {\bibfnamefont {Y.~S.}\ \bibnamefont
  {Kivshar}},\ and\ \bibinfo {author} {\bibfnamefont {B.}~\bibnamefont
  {Luk'yanchuk}},\ }\bibfield  {title} {\bibinfo {title} {{Optically resonant
  dielectric nanostructures}},\ }\href
  {https://doi.org/10.1126/science.aag2472} {\bibfield  {journal} {\bibinfo
  {journal} {Science}\ }\textbf {\bibinfo {volume} {354}},\ \bibinfo {pages}
  {aag2472} (\bibinfo {year} {2016})}\BibitemShut {NoStop}%
\bibitem [{\citenamefont {Kruk}\ and\ \citenamefont
  {Kivshar}(2017)}]{Kruk-Kivshar}%
  \BibitemOpen
  \bibfield  {author} {\bibinfo {author} {\bibfnamefont {S.}~\bibnamefont
  {Kruk}}\ and\ \bibinfo {author} {\bibfnamefont {Y.}~\bibnamefont {Kivshar}},\
  }\bibfield  {title} {\bibinfo {title} {{Functional Meta-Optics and
  Nanophotonics Governed by Mie Resonances}},\ }\href
  {https://doi.org/10.1021/acsphotonics.7b01038} {\bibfield  {journal}
  {\bibinfo  {journal} {ACS Photonics}\ }\textbf {\bibinfo {volume} {4}},\
  \bibinfo {pages} {2638} (\bibinfo {year} {2017})}\BibitemShut {NoStop}%
\bibitem [{\citenamefont {Staude}\ \emph {et~al.}(2013)\citenamefont {Staude},
  \citenamefont {Miroshnichenko}, \citenamefont {Decker}, \citenamefont
  {Fofang}, \citenamefont {Liu}, \citenamefont {Gonzales}, \citenamefont
  {Dominguez}, \citenamefont {Luk}, \citenamefont {Neshev}, \citenamefont
  {Brener},\ and\ \citenamefont {Kivshar}}]{Staude-13}%
  \BibitemOpen
  \bibfield  {author} {\bibinfo {author} {\bibfnamefont {I.}~\bibnamefont
  {Staude}}, \bibinfo {author} {\bibfnamefont {A.~E.}\ \bibnamefont
  {Miroshnichenko}}, \bibinfo {author} {\bibfnamefont {M.}~\bibnamefont
  {Decker}}, \bibinfo {author} {\bibfnamefont {N.~T.}\ \bibnamefont {Fofang}},
  \bibinfo {author} {\bibfnamefont {S.}~\bibnamefont {Liu}}, \bibinfo {author}
  {\bibfnamefont {E.}~\bibnamefont {Gonzales}}, \bibinfo {author}
  {\bibfnamefont {J.}~\bibnamefont {Dominguez}}, \bibinfo {author}
  {\bibfnamefont {T.~S.}\ \bibnamefont {Luk}}, \bibinfo {author} {\bibfnamefont
  {D.~N.}\ \bibnamefont {Neshev}}, \bibinfo {author} {\bibfnamefont
  {I.}~\bibnamefont {Brener}},\ and\ \bibinfo {author} {\bibfnamefont
  {Y.}~\bibnamefont {Kivshar}},\ }\bibfield  {title} {\bibinfo {title}
  {{Tailoring Directional Scattering through Magnetic and Electric Resonances
  in Subwavelength Silicon Nanodisks}},\ }\href
  {https://doi.org/10.1021/nn402736f} {\bibfield  {journal} {\bibinfo
  {journal} {ACS Nano}\ }\textbf {\bibinfo {volume} {7}},\ \bibinfo {pages}
  {7824} (\bibinfo {year} {2013})}\BibitemShut {NoStop}%
\bibitem [{\citenamefont {Decker}\ \emph {et~al.}(2015)\citenamefont {Decker},
  \citenamefont {Staude}, \citenamefont {Falkner}, \citenamefont {Dominguez},
  \citenamefont {Neshev}, \citenamefont {Brener}, \citenamefont {Pertsch},\
  and\ \citenamefont {Kivshar}}]{Decker}%
  \BibitemOpen
  \bibfield  {author} {\bibinfo {author} {\bibfnamefont {M.}~\bibnamefont
  {Decker}}, \bibinfo {author} {\bibfnamefont {I.}~\bibnamefont {Staude}},
  \bibinfo {author} {\bibfnamefont {M.}~\bibnamefont {Falkner}}, \bibinfo
  {author} {\bibfnamefont {J.}~\bibnamefont {Dominguez}}, \bibinfo {author}
  {\bibfnamefont {D.~N.}\ \bibnamefont {Neshev}}, \bibinfo {author}
  {\bibfnamefont {I.}~\bibnamefont {Brener}}, \bibinfo {author} {\bibfnamefont
  {T.}~\bibnamefont {Pertsch}},\ and\ \bibinfo {author} {\bibfnamefont {Y.~S.}\
  \bibnamefont {Kivshar}},\ }\bibfield  {title} {\bibinfo {title}
  {{High-Efficiency Dielectric Huygens' Surfaces}},\ }\href
  {https://doi.org/10.1002/adom.201400584} {\bibfield  {journal} {\bibinfo
  {journal} {Adv. Opt. Mater.}\ }\textbf {\bibinfo {volume} {3}},\ \bibinfo
  {pages} {813} (\bibinfo {year} {2015})}\BibitemShut {NoStop}%
\bibitem [{\citenamefont {Kruk}\ \emph {et~al.}(2016)\citenamefont {Kruk},
  \citenamefont {Hopkins}, \citenamefont {Kravchenko}, \citenamefont
  {Miroshnichenko}, \citenamefont {Neshev},\ and\ \citenamefont
  {Kivshar}}]{Kruk-APL}%
  \BibitemOpen
  \bibfield  {author} {\bibinfo {author} {\bibfnamefont {S.}~\bibnamefont
  {Kruk}}, \bibinfo {author} {\bibfnamefont {B.}~\bibnamefont {Hopkins}},
  \bibinfo {author} {\bibfnamefont {I.~I.}\ \bibnamefont {Kravchenko}},
  \bibinfo {author} {\bibfnamefont {A.}~\bibnamefont {Miroshnichenko}},
  \bibinfo {author} {\bibfnamefont {D.~N.}\ \bibnamefont {Neshev}},\ and\
  \bibinfo {author} {\bibfnamefont {Y.~S.}\ \bibnamefont {Kivshar}},\
  }\bibfield  {title} {\bibinfo {title} {{Broadband highly efficient dielectric
  metadevices for polarization control}},\ }\href
  {https://doi.org/10.1063/1.4949007} {\bibfield  {journal} {\bibinfo
  {journal} {APL Photonics}\ }\textbf {\bibinfo {volume} {1}},\ \bibinfo
  {pages} {030801} (\bibinfo {year} {2016})}\BibitemShut {NoStop}%
\bibitem [{\citenamefont {Rybin}\ \emph {et~al.}(2017)\citenamefont {Rybin},
  \citenamefont {Koshelev}, \citenamefont {Sadrieva}, \citenamefont {Samusev},
  \citenamefont {Bogdanov}, \citenamefont {Limonov},\ and\ \citenamefont
  {Kivshar}}]{Koshelev}%
  \BibitemOpen
  \bibfield  {author} {\bibinfo {author} {\bibfnamefont {M.~V.}\ \bibnamefont
  {Rybin}}, \bibinfo {author} {\bibfnamefont {K.~L.}\ \bibnamefont {Koshelev}},
  \bibinfo {author} {\bibfnamefont {Z.~F.}\ \bibnamefont {Sadrieva}}, \bibinfo
  {author} {\bibfnamefont {K.~B.}\ \bibnamefont {Samusev}}, \bibinfo {author}
  {\bibfnamefont {A.~A.}\ \bibnamefont {Bogdanov}}, \bibinfo {author}
  {\bibfnamefont {M.~F.}\ \bibnamefont {Limonov}},\ and\ \bibinfo {author}
  {\bibfnamefont {Y.~S.}\ \bibnamefont {Kivshar}},\ }\bibfield  {title}
  {\bibinfo {title} {{High-$Q$ Supercavity Modes in Subwavelength Dielectric
  Resonators}},\ }\href {https://doi.org/10.1103/PhysRevLett.119.243901}
  {\bibfield  {journal} {\bibinfo  {journal} {Phys. Rev. Lett.}\ }\textbf
  {\bibinfo {volume} {119}},\ \bibinfo {pages} {243901} (\bibinfo {year}
  {2017})}\BibitemShut {NoStop}%
\bibitem [{\citenamefont {Shcherbakov}\ \emph {et~al.}(2014)\citenamefont
  {Shcherbakov}, \citenamefont {Neshev}, \citenamefont {Hopkins}, \citenamefont
  {Shorokhov}, \citenamefont {Staude}, \citenamefont {Melik-Gaykazyan},
  \citenamefont {Decker}, \citenamefont {Ezhov}, \citenamefont
  {Miroshnichenko}, \citenamefont {Brener}, \citenamefont {Fedyanin},\ and\
  \citenamefont {Kivshar}}]{Shcherbakov}%
  \BibitemOpen
  \bibfield  {author} {\bibinfo {author} {\bibfnamefont {M.~R.}\ \bibnamefont
  {Shcherbakov}}, \bibinfo {author} {\bibfnamefont {D.~N.}\ \bibnamefont
  {Neshev}}, \bibinfo {author} {\bibfnamefont {B.}~\bibnamefont {Hopkins}},
  \bibinfo {author} {\bibfnamefont {A.~S.}\ \bibnamefont {Shorokhov}}, \bibinfo
  {author} {\bibfnamefont {I.}~\bibnamefont {Staude}}, \bibinfo {author}
  {\bibfnamefont {E.~V.}\ \bibnamefont {Melik-Gaykazyan}}, \bibinfo {author}
  {\bibfnamefont {M.}~\bibnamefont {Decker}}, \bibinfo {author} {\bibfnamefont
  {A.~A.}\ \bibnamefont {Ezhov}}, \bibinfo {author} {\bibfnamefont {A.~E.}\
  \bibnamefont {Miroshnichenko}}, \bibinfo {author} {\bibfnamefont
  {I.}~\bibnamefont {Brener}}, \bibinfo {author} {\bibfnamefont {A.~A.}\
  \bibnamefont {Fedyanin}},\ and\ \bibinfo {author} {\bibfnamefont {Y.~S.}\
  \bibnamefont {Kivshar}},\ }\bibfield  {title} {\bibinfo {title} {{Enhanced
  Third-Harmonic Generation in Silicon Nanoparticles Driven by Magnetic
  Response}},\ }\href {https://doi.org/10.1021/nl503029j} {\bibfield  {journal}
  {\bibinfo  {journal} {Nano Lett.}\ }\textbf {\bibinfo {volume} {14}},\
  \bibinfo {pages} {6488} (\bibinfo {year} {2014})}\BibitemShut {NoStop}%
\bibitem [{\citenamefont {Koshelev}\ \emph {et~al.}(2020)\citenamefont
  {Koshelev}, \citenamefont {Kruk}, \citenamefont {Melik-Gaykazyan},
  \citenamefont {Choi}, \citenamefont {Bogdanov}, \citenamefont {Park},\ and\
  \citenamefont {Kivshar}}]{Koshelev-Science}%
  \BibitemOpen
  \bibfield  {author} {\bibinfo {author} {\bibfnamefont {K.}~\bibnamefont
  {Koshelev}}, \bibinfo {author} {\bibfnamefont {S.}~\bibnamefont {Kruk}},
  \bibinfo {author} {\bibfnamefont {E.}~\bibnamefont {Melik-Gaykazyan}},
  \bibinfo {author} {\bibfnamefont {J.-H.}\ \bibnamefont {Choi}}, \bibinfo
  {author} {\bibfnamefont {A.}~\bibnamefont {Bogdanov}}, \bibinfo {author}
  {\bibfnamefont {H.-G.}\ \bibnamefont {Park}},\ and\ \bibinfo {author}
  {\bibfnamefont {Y.}~\bibnamefont {Kivshar}},\ }\bibfield  {title} {\bibinfo
  {title} {{Subwavelength dielectric resonators for nonlinear nanophotonics}},\
  }\href {https://doi.org/10.1126/science.aaz3985} {\bibfield  {journal}
  {\bibinfo  {journal} {Science}\ }\textbf {\bibinfo {volume} {367}},\ \bibinfo
  {pages} {288} (\bibinfo {year} {2020})}\BibitemShut {NoStop}%
\bibitem [{\citenamefont {Tittl}\ \emph {et~al.}(2018)\citenamefont {Tittl},
  \citenamefont {Leitis}, \citenamefont {Liu}, \citenamefont {Yesilkoy},
  \citenamefont {Choi}, \citenamefont {Neshev}, \citenamefont {Kivshar},\ and\
  \citenamefont {Altug}}]{Tittl}%
  \BibitemOpen
  \bibfield  {author} {\bibinfo {author} {\bibfnamefont {A.}~\bibnamefont
  {Tittl}}, \bibinfo {author} {\bibfnamefont {A.}~\bibnamefont {Leitis}},
  \bibinfo {author} {\bibfnamefont {M.}~\bibnamefont {Liu}}, \bibinfo {author}
  {\bibfnamefont {F.}~\bibnamefont {Yesilkoy}}, \bibinfo {author}
  {\bibfnamefont {D.-Y.}\ \bibnamefont {Choi}}, \bibinfo {author}
  {\bibfnamefont {D.~N.}\ \bibnamefont {Neshev}}, \bibinfo {author}
  {\bibfnamefont {Y.~S.}\ \bibnamefont {Kivshar}},\ and\ \bibinfo {author}
  {\bibfnamefont {H.}~\bibnamefont {Altug}},\ }\bibfield  {title} {\bibinfo
  {title} {{Imaging-based molecular barcoding with pixelated dielectric
  metasurfaces}},\ }\href {https://doi.org/10.1126/science.aas9768} {\bibfield
  {journal} {\bibinfo  {journal} {Science}\ }\textbf {\bibinfo {volume}
  {360}},\ \bibinfo {pages} {1105} (\bibinfo {year} {2018})}\BibitemShut
  {NoStop}%
\bibitem [{\citenamefont {Evlyukhin}\ \emph {et~al.}(2012)\citenamefont
  {Evlyukhin}, \citenamefont {Novikov}, \citenamefont {Zywietz}, \citenamefont
  {Eriksen}, \citenamefont {Reinhardt}, \citenamefont {Bozhevolnyi},\ and\
  \citenamefont {Chichkov}}]{Evlyukhin-NL}%
  \BibitemOpen
  \bibfield  {author} {\bibinfo {author} {\bibfnamefont {A.~B.}\ \bibnamefont
  {Evlyukhin}}, \bibinfo {author} {\bibfnamefont {S.~M.}\ \bibnamefont
  {Novikov}}, \bibinfo {author} {\bibfnamefont {U.}~\bibnamefont {Zywietz}},
  \bibinfo {author} {\bibfnamefont {R.~L.}\ \bibnamefont {Eriksen}}, \bibinfo
  {author} {\bibfnamefont {C.}~\bibnamefont {Reinhardt}}, \bibinfo {author}
  {\bibfnamefont {S.~I.}\ \bibnamefont {Bozhevolnyi}},\ and\ \bibinfo {author}
  {\bibfnamefont {B.~N.}\ \bibnamefont {Chichkov}},\ }\bibfield  {title}
  {\bibinfo {title} {{Demonstration of Magnetic Dipole Resonances of Dielectric
  Nanospheres in the Visible Region}},\ }\href
  {https://doi.org/10.1021/nl301594s} {\bibfield  {journal} {\bibinfo
  {journal} {Nano Lett.}\ }\textbf {\bibinfo {volume} {12}},\ \bibinfo {pages}
  {3749} (\bibinfo {year} {2012})}\BibitemShut {NoStop}%
\bibitem [{\citenamefont {Kuznetsov}\ \emph {et~al.}(2012)\citenamefont
  {Kuznetsov}, \citenamefont {Miroshnichenko}, \citenamefont {Fu},
  \citenamefont {Zhang},\ and\ \citenamefont {Luk'yanchuk}}]{Kuznetsov-SciRep}%
  \BibitemOpen
  \bibfield  {author} {\bibinfo {author} {\bibfnamefont {A.~I.}\ \bibnamefont
  {Kuznetsov}}, \bibinfo {author} {\bibfnamefont {A.~E.}\ \bibnamefont
  {Miroshnichenko}}, \bibinfo {author} {\bibfnamefont {Y.~H.}\ \bibnamefont
  {Fu}}, \bibinfo {author} {\bibfnamefont {J.}~\bibnamefont {Zhang}},\ and\
  \bibinfo {author} {\bibfnamefont {B.}~\bibnamefont {Luk'yanchuk}},\
  }\bibfield  {title} {\bibinfo {title} {Magnetic light},\ }\href
  {https://doi.org/10.1038/srep00492} {\bibfield  {journal} {\bibinfo
  {journal} {Sci. Rep.}\ }\textbf {\bibinfo {volume} {2}},\ \bibinfo {pages}
  {492} (\bibinfo {year} {2012})}\BibitemShut {NoStop}%
\bibitem [{\citenamefont {Smirnova}\ and\ \citenamefont
  {Kivshar}(2016)}]{Smirnova-Optica}%
  \BibitemOpen
  \bibfield  {author} {\bibinfo {author} {\bibfnamefont {D.}~\bibnamefont
  {Smirnova}}\ and\ \bibinfo {author} {\bibfnamefont {Y.~S.}\ \bibnamefont
  {Kivshar}},\ }\bibfield  {title} {\bibinfo {title} {Multipolar nonlinear
  nanophotonics},\ }\href {https://doi.org/10.1364/OPTICA.3.001241} {\bibfield
  {journal} {\bibinfo  {journal} {Optica}\ }\textbf {\bibinfo {volume} {3}},\
  \bibinfo {pages} {1241} (\bibinfo {year} {2016})}\BibitemShut {NoStop}%
\bibitem [{\citenamefont {Jackson}(1998)}]{Jackson}%
  \BibitemOpen
  \bibfield  {author} {\bibinfo {author} {\bibfnamefont {J.~D.}\ \bibnamefont
  {Jackson}},\ }\href@noop {} {\emph {\bibinfo {title} {{Classical
  Electrodynamics}}}}\ (\bibinfo  {publisher} {Wiley, New York},\ \bibinfo
  {year} {1998})\BibitemShut {NoStop}%
\bibitem [{\citenamefont {Purcell}\ and\ \citenamefont
  {Pennypacker}(1973)}]{Purcell}%
  \BibitemOpen
  \bibfield  {author} {\bibinfo {author} {\bibfnamefont {E.~M.}\ \bibnamefont
  {Purcell}}\ and\ \bibinfo {author} {\bibfnamefont {C.~R.}\ \bibnamefont
  {Pennypacker}},\ }\bibfield  {title} {\bibinfo {title} {Scattering and
  absorption of light by nonspherical dielectric grains},\ }\href@noop {}
  {\bibfield  {journal} {\bibinfo  {journal} {The Astrophysical Journal}\
  }\textbf {\bibinfo {volume} {186}},\ \bibinfo {pages} {705} (\bibinfo {year}
  {1973})}\BibitemShut {NoStop}%
\bibitem [{\citenamefont {Draine}(1988)}]{Draine}%
  \BibitemOpen
  \bibfield  {author} {\bibinfo {author} {\bibfnamefont {B.~T.}\ \bibnamefont
  {Draine}},\ }\bibfield  {title} {\bibinfo {title} {The discrete-dipole
  approximation and its application to interstellar graphite grains},\
  }\href@noop {} {\bibfield  {journal} {\bibinfo  {journal} {The Astrophysical
  Journal}\ }\textbf {\bibinfo {volume} {333}},\ \bibinfo {pages} {848}
  (\bibinfo {year} {1988})}\BibitemShut {NoStop}%
\bibitem [{\citenamefont {Yurkin}\ and\ \citenamefont
  {Hoekstra}(2007)}]{Yurkin}%
  \BibitemOpen
  \bibfield  {author} {\bibinfo {author} {\bibfnamefont {M.~A.}\ \bibnamefont
  {Yurkin}}\ and\ \bibinfo {author} {\bibfnamefont {A.~G.}\ \bibnamefont
  {Hoekstra}},\ }\bibfield  {title} {\bibinfo {title} {{The discrete dipole
  approximation: An overview and recent developments}},\ }\href
  {https://doi.org/https://doi.org/10.1016/j.jqsrt.2007.01.034} {\bibfield
  {journal} {\bibinfo  {journal} {J. Quant. Spectrosc. Radiat. Transfer}\
  }\textbf {\bibinfo {volume} {106}},\ \bibinfo {pages} {558} (\bibinfo {year}
  {2007})}\BibitemShut {NoStop}%
\bibitem [{\citenamefont {Evlyukhin}\ \emph {et~al.}(2011)\citenamefont
  {Evlyukhin}, \citenamefont {Reinhardt},\ and\ \citenamefont
  {Chichkov}}]{Evlyukhin-11}%
  \BibitemOpen
  \bibfield  {author} {\bibinfo {author} {\bibfnamefont {A.~B.}\ \bibnamefont
  {Evlyukhin}}, \bibinfo {author} {\bibfnamefont {C.}~\bibnamefont
  {Reinhardt}},\ and\ \bibinfo {author} {\bibfnamefont {B.~N.}\ \bibnamefont
  {Chichkov}},\ }\bibfield  {title} {\bibinfo {title} {{Multipole light
  scattering by nonspherical nanoparticles in the discrete dipole
  approximation}},\ }\href {https://doi.org/10.1103/PhysRevB.84.235429}
  {\bibfield  {journal} {\bibinfo  {journal} {Phys. Rev. B}\ }\textbf {\bibinfo
  {volume} {84}},\ \bibinfo {pages} {235429} (\bibinfo {year}
  {2011})}\BibitemShut {NoStop}%
\bibitem [{\citenamefont {Baryshnikova}\ \emph {et~al.}(2019)\citenamefont
  {Baryshnikova}, \citenamefont {Smirnova}, \citenamefont {Luk'yanchuk},\ and\
  \citenamefont {Kivshar}}]{Baryshnikova-adom}%
  \BibitemOpen
  \bibfield  {author} {\bibinfo {author} {\bibfnamefont {K.~V.}\ \bibnamefont
  {Baryshnikova}}, \bibinfo {author} {\bibfnamefont {D.~A.}\ \bibnamefont
  {Smirnova}}, \bibinfo {author} {\bibfnamefont {B.~S.}\ \bibnamefont
  {Luk'yanchuk}},\ and\ \bibinfo {author} {\bibfnamefont {Y.~S.}\ \bibnamefont
  {Kivshar}},\ }\bibfield  {title} {\bibinfo {title} {{Optical Anapoles:
  Concepts and Applications}},\ }\href {https://doi.org/10.1002/adom.201801350}
  {\bibfield  {journal} {\bibinfo  {journal} {Adv. Opt. Mater.}\ }\textbf
  {\bibinfo {volume} {7}},\ \bibinfo {pages} {1801350} (\bibinfo {year}
  {2019})}\BibitemShut {NoStop}%
\bibitem [{\citenamefont {Raza}\ \emph {et~al.}(2015)\citenamefont {Raza},
  \citenamefont {Bozhevolnyi}, \citenamefont {Wubs},\ and\ \citenamefont
  {Mortensen}}]{Raza}%
  \BibitemOpen
  \bibfield  {author} {\bibinfo {author} {\bibfnamefont {S.}~\bibnamefont
  {Raza}}, \bibinfo {author} {\bibfnamefont {S.~I.}\ \bibnamefont
  {Bozhevolnyi}}, \bibinfo {author} {\bibfnamefont {M.}~\bibnamefont {Wubs}},\
  and\ \bibinfo {author} {\bibfnamefont {N.~A.}\ \bibnamefont {Mortensen}},\
  }\bibfield  {title} {\bibinfo {title} {{Nonlocal optical response in metallic
  nanostructures}},\ }\href
  {https://doi.org/https://doi.org/10.1088/0953-8984/27/18/183204} {\bibfield
  {journal} {\bibinfo  {journal} {J. Phys.: Condens. Matter}\ }\textbf
  {\bibinfo {volume} {27}},\ \bibinfo {pages} {183204} (\bibinfo {year}
  {2015})}\BibitemShut {NoStop}%
\bibitem [{\citenamefont {David}\ and\ \citenamefont {de~Abajo}(2011)}]{Abajo}%
  \BibitemOpen
  \bibfield  {author} {\bibinfo {author} {\bibfnamefont {C.}~\bibnamefont
  {David}}\ and\ \bibinfo {author} {\bibfnamefont {F.~J.~G.}\ \bibnamefont
  {de~Abajo}},\ }\bibfield  {title} {\bibinfo {title} {{Spatial Nonlocality in
  the Optical Response of Metal Nanoparticles}},\ }\href
  {https://doi.org/https://doi.org/10.1021/jp204261u} {\bibfield  {journal}
  {\bibinfo  {journal} {J. Phys. Chem. C}\ }\textbf {\bibinfo {volume} {115}},\
  \bibinfo {pages} {19470} (\bibinfo {year} {2011})}\BibitemShut {NoStop}%
\bibitem [{\citenamefont {Fernandez-Corbaton}\ \emph
  {et~al.}(2015)\citenamefont {Fernandez-Corbaton}, \citenamefont {Nanz},
  \citenamefont {Alaee},\ and\ \citenamefont {Rockstuhl}}]{Rockstuhl-OE-15}%
  \BibitemOpen
  \bibfield  {author} {\bibinfo {author} {\bibfnamefont {I.}~\bibnamefont
  {Fernandez-Corbaton}}, \bibinfo {author} {\bibfnamefont {S.}~\bibnamefont
  {Nanz}}, \bibinfo {author} {\bibfnamefont {R.}~\bibnamefont {Alaee}},\ and\
  \bibinfo {author} {\bibfnamefont {C.}~\bibnamefont {Rockstuhl}},\ }\bibfield
  {title} {\bibinfo {title} {{Exact dipolar moments of a localized electric
  current distribution}},\ }\href
  {https://doi.org/https://doi.org/10.1364/OE.23.033044} {\bibfield  {journal}
  {\bibinfo  {journal} {Opt. Express}\ }\textbf {\bibinfo {volume} {23}},\
  \bibinfo {pages} {33044} (\bibinfo {year} {2015})}\BibitemShut {NoStop}%
\bibitem [{\citenamefont {Alaee}\ \emph {et~al.}(2018)\citenamefont {Alaee},
  \citenamefont {Rockstuhl},\ and\ \citenamefont
  {Fernandez-Corbaton}}]{Rockstuhl-OC-18}%
  \BibitemOpen
  \bibfield  {author} {\bibinfo {author} {\bibfnamefont {R.}~\bibnamefont
  {Alaee}}, \bibinfo {author} {\bibfnamefont {C.}~\bibnamefont {Rockstuhl}},\
  and\ \bibinfo {author} {\bibfnamefont {I.}~\bibnamefont
  {Fernandez-Corbaton}},\ }\bibfield  {title} {\bibinfo {title} {{An
  electromagnetic multipole expansion beyond the long-wavelength
  approximation}},\ }\href
  {https://doi.org/https://doi.org/10.1016/j.optcom.2017.08.064} {\bibfield
  {journal} {\bibinfo  {journal} {Opt. Commun.}\ }\textbf {\bibinfo {volume}
  {407}},\ \bibinfo {pages} {17} (\bibinfo {year} {2018})}\BibitemShut
  {NoStop}%
\bibitem [{\citenamefont {Evlyukhin}\ \emph {et~al.}(2016)\citenamefont
  {Evlyukhin}, \citenamefont {Fischer}, \citenamefont {Reinhardt},\ and\
  \citenamefont {Chichkov}}]{Evlyukhin-16}%
  \BibitemOpen
  \bibfield  {author} {\bibinfo {author} {\bibfnamefont {A.~B.}\ \bibnamefont
  {Evlyukhin}}, \bibinfo {author} {\bibfnamefont {T.}~\bibnamefont {Fischer}},
  \bibinfo {author} {\bibfnamefont {C.}~\bibnamefont {Reinhardt}},\ and\
  \bibinfo {author} {\bibfnamefont {B.~N.}\ \bibnamefont {Chichkov}},\
  }\bibfield  {title} {\bibinfo {title} {{Optical theorem and multipole
  scattering of light by arbitrarily shaped nanoparticles}},\ }\href
  {https://doi.org/10.1103/PhysRevB.94.205434} {\bibfield  {journal} {\bibinfo
  {journal} {Phys. Rev. B}\ }\textbf {\bibinfo {volume} {94}},\ \bibinfo
  {pages} {205434} (\bibinfo {year} {2016})}\BibitemShut {NoStop}%
\bibitem [{\citenamefont {Gurvitz}\ \emph {et~al.}(2019)\citenamefont
  {Gurvitz}, \citenamefont {Ladutenko}, \citenamefont {Dergachev},
  \citenamefont {Evlyukhin}, \citenamefont {Miroshnichenko},\ and\
  \citenamefont {Shalin}}]{Gurvitz}%
  \BibitemOpen
  \bibfield  {author} {\bibinfo {author} {\bibfnamefont {E.~A.}\ \bibnamefont
  {Gurvitz}}, \bibinfo {author} {\bibfnamefont {K.~S.}\ \bibnamefont
  {Ladutenko}}, \bibinfo {author} {\bibfnamefont {P.~A.}\ \bibnamefont
  {Dergachev}}, \bibinfo {author} {\bibfnamefont {A.~B.}\ \bibnamefont
  {Evlyukhin}}, \bibinfo {author} {\bibfnamefont {A.~E.}\ \bibnamefont
  {Miroshnichenko}},\ and\ \bibinfo {author} {\bibfnamefont {A.~S.}\
  \bibnamefont {Shalin}},\ }\bibfield  {title} {\bibinfo {title} {{The
  High-Order Toroidal Moments and Anapole States in All-Dielectric
  Photonics}},\ }\href {https://doi.org/10.1002/lpor.201800266} {\bibfield
  {journal} {\bibinfo  {journal} {Laser Photonics Rev.}\ }\textbf {\bibinfo
  {volume} {13}},\ \bibinfo {pages} {1800266} (\bibinfo {year}
  {2019})}\BibitemShut {NoStop}%
\bibitem [{\citenamefont {Landau}\ and\ \citenamefont
  {Lifshitz}(1980)}]{landau5}%
  \BibitemOpen
  \bibfield  {author} {\bibinfo {author} {\bibfnamefont {L.~D.}\ \bibnamefont
  {Landau}}\ and\ \bibinfo {author} {\bibfnamefont {E.~M.}\ \bibnamefont
  {Lifshitz}},\ }\href@noop {} {\emph {\bibinfo {title} {Statistical Physics,
  Part 1}}}\ (\bibinfo  {publisher} {Pergamon Press, Oxford},\ \bibinfo {year}
  {1980})\BibitemShut {NoStop}%
\bibitem [{\citenamefont {Gorlach}\ \emph {et~al.}(2019)\citenamefont
  {Gorlach}, \citenamefont {Zhirihin}, \citenamefont {Slobozhanyuk},
  \citenamefont {Khanikaev},\ and\ \citenamefont {Gorlach}}]{GorlachPRB2019}%
  \BibitemOpen
  \bibfield  {author} {\bibinfo {author} {\bibfnamefont {A.~A.}\ \bibnamefont
  {Gorlach}}, \bibinfo {author} {\bibfnamefont {D.~V.}\ \bibnamefont
  {Zhirihin}}, \bibinfo {author} {\bibfnamefont {A.~P.}\ \bibnamefont
  {Slobozhanyuk}}, \bibinfo {author} {\bibfnamefont {A.~B.}\ \bibnamefont
  {Khanikaev}},\ and\ \bibinfo {author} {\bibfnamefont {M.~A.}\ \bibnamefont
  {Gorlach}},\ }\bibfield  {title} {\bibinfo {title} {{Photonic Jackiw-Rebbi
  states in all-dielectric structures controlled by bianisotropy}},\ }\href
  {https://doi.org/10.1103/physrevb.99.205122} {\bibfield  {journal} {\bibinfo
  {journal} {Phys. Rev. B}\ }\textbf {\bibinfo {volume} {99}},\ \bibinfo
  {pages} {205122} (\bibinfo {year} {2019})}\BibitemShut {NoStop}%
\bibitem [{\citenamefont {Belov}\ and\ \citenamefont
  {Simovski}(2006)}]{Belov2006}%
  \BibitemOpen
  \bibfield  {author} {\bibinfo {author} {\bibfnamefont {P.~A.}\ \bibnamefont
  {Belov}}\ and\ \bibinfo {author} {\bibfnamefont {C.~R.}\ \bibnamefont
  {Simovski}},\ }\bibfield  {title} {\bibinfo {title} {{Boundary conditions for
  interfaces of electromagnetic crystals and the generalized Ewald-Oseen
  extinction principle}},\ }\href {https://doi.org/10.1103/physrevb.73.045102}
  {\bibfield  {journal} {\bibinfo  {journal} {Phys. Rev. B}\ }\textbf {\bibinfo
  {volume} {73}},\ \bibinfo {pages} {045102} (\bibinfo {year}
  {2006})}\BibitemShut {NoStop}%
\bibitem [{\citenamefont {Chebykin}\ \emph {et~al.}(2015)\citenamefont
  {Chebykin}, \citenamefont {Gorlach},\ and\ \citenamefont {Belov}}]{Chebykin}%
  \BibitemOpen
  \bibfield  {author} {\bibinfo {author} {\bibfnamefont {A.~V.}\ \bibnamefont
  {Chebykin}}, \bibinfo {author} {\bibfnamefont {M.~A.}\ \bibnamefont
  {Gorlach}},\ and\ \bibinfo {author} {\bibfnamefont {P.~A.}\ \bibnamefont
  {Belov}},\ }\bibfield  {title} {\bibinfo {title} {Spatial-dispersion-induced
  birefringence in metamaterials with cubic symmetry},\ }\href
  {https://doi.org/10.1103/PhysRevB.92.045127} {\bibfield  {journal} {\bibinfo
  {journal} {Phys. Rev. B}\ }\textbf {\bibinfo {volume} {92}},\ \bibinfo
  {pages} {045127} (\bibinfo {year} {2015})}\BibitemShut {NoStop}%
\bibitem [{\citenamefont {Agranovich}\ and\ \citenamefont
  {Ginzburg}(1984)}]{Agranovich}%
  \BibitemOpen
  \bibfield  {author} {\bibinfo {author} {\bibfnamefont {V.~M.}\ \bibnamefont
  {Agranovich}}\ and\ \bibinfo {author} {\bibfnamefont {V.~L.}\ \bibnamefont
  {Ginzburg}},\ }\href@noop {} {\emph {\bibinfo {title} {{Crystal Optics with
  Spatial Dispersion and Excitons}}}}\ (\bibinfo  {publisher} {Springer,
  Berlin},\ \bibinfo {year} {1984})\BibitemShut {NoStop}%
\bibitem [{\citenamefont {Silveirinha}(2007{\natexlab{a}})}]{Silveirinha2007}%
  \BibitemOpen
  \bibfield  {author} {\bibinfo {author} {\bibfnamefont {M.~G.}\ \bibnamefont
  {Silveirinha}},\ }\bibfield  {title} {\bibinfo {title} {Metamaterial
  homogenization approach with application to the characterization of
  microstructured composites with negative parameters},\ }\href
  {https://doi.org/10.1103/physrevb.75.115104} {\bibfield  {journal} {\bibinfo
  {journal} {Phys. Rev. B}\ }\textbf {\bibinfo {volume} {75}},\ \bibinfo
  {pages} {115104} (\bibinfo {year} {2007}{\natexlab{a}})}\BibitemShut
  {NoStop}%
\bibitem [{\citenamefont {Mnasri}\ \emph {et~al.}(2018)\citenamefont {Mnasri},
  \citenamefont {Khrabustovskyi}, \citenamefont {Stohrer}, \citenamefont
  {Plum},\ and\ \citenamefont {Rockstuhl}}]{Mnasri2018}%
  \BibitemOpen
  \bibfield  {author} {\bibinfo {author} {\bibfnamefont {K.}~\bibnamefont
  {Mnasri}}, \bibinfo {author} {\bibfnamefont {A.}~\bibnamefont
  {Khrabustovskyi}}, \bibinfo {author} {\bibfnamefont {C.}~\bibnamefont
  {Stohrer}}, \bibinfo {author} {\bibfnamefont {M.}~\bibnamefont {Plum}},\ and\
  \bibinfo {author} {\bibfnamefont {C.}~\bibnamefont {Rockstuhl}},\ }\bibfield
  {title} {\bibinfo {title} {Beyond local effective material properties for
  metamaterials},\ }\href {https://doi.org/10.1103/physrevb.97.075439}
  {\bibfield  {journal} {\bibinfo  {journal} {Phys. Rev. B}\ }\textbf {\bibinfo
  {volume} {97}},\ \bibinfo {pages} {075439} (\bibinfo {year}
  {2018})}\BibitemShut {NoStop}%
\bibitem [{\citenamefont {Silveirinha}(2007{\natexlab{b}})}]{Silv-L-2007}%
  \BibitemOpen
  \bibfield  {author} {\bibinfo {author} {\bibfnamefont {M.~G.}\ \bibnamefont
  {Silveirinha}},\ }\bibfield  {title} {\bibinfo {title} {{Generalized
  Lorentz-Lorenz formulas for microstructured materials}},\ }\href
  {https://doi.org/10.1103/PhysRevB.76.245117} {\bibfield  {journal} {\bibinfo
  {journal} {Phys. Rev. B}\ }\textbf {\bibinfo {volume} {76}},\ \bibinfo
  {pages} {245117} (\bibinfo {year} {2007}{\natexlab{b}})}\BibitemShut
  {NoStop}%
\bibitem [{\citenamefont {Gorlach}\ and\ \citenamefont
  {Belov}(2014)}]{Gorlach-14}%
  \BibitemOpen
  \bibfield  {author} {\bibinfo {author} {\bibfnamefont {M.~A.}\ \bibnamefont
  {Gorlach}}\ and\ \bibinfo {author} {\bibfnamefont {P.~A.}\ \bibnamefont
  {Belov}},\ }\bibfield  {title} {\bibinfo {title} {{Effect of spatial
  dispersion on the topological transition in metamaterials}},\ }\href
  {https://doi.org/10.1103/PhysRevB.90.115136} {\bibfield  {journal} {\bibinfo
  {journal} {Phys. Rev. B}\ }\textbf {\bibinfo {volume} {90}},\ \bibinfo
  {pages} {115136} (\bibinfo {year} {2014})}\BibitemShut {NoStop}%
\bibitem [{\citenamefont {Patoux}\ \emph {et~al.}(2020)\citenamefont {Patoux},
  \citenamefont {Majorel}, \citenamefont {Wiecha}, \citenamefont {Cuche},
  \citenamefont {Muskens}, \citenamefont {Girard},\ and\ \citenamefont
  {Arbouet}}]{Arbouet}%
  \BibitemOpen
  \bibfield  {author} {\bibinfo {author} {\bibfnamefont {A.}~\bibnamefont
  {Patoux}}, \bibinfo {author} {\bibfnamefont {C.}~\bibnamefont {Majorel}},
  \bibinfo {author} {\bibfnamefont {P.~R.}\ \bibnamefont {Wiecha}}, \bibinfo
  {author} {\bibfnamefont {A.}~\bibnamefont {Cuche}}, \bibinfo {author}
  {\bibfnamefont {O.~L.}\ \bibnamefont {Muskens}}, \bibinfo {author}
  {\bibfnamefont {C.}~\bibnamefont {Girard}},\ and\ \bibinfo {author}
  {\bibfnamefont {A.}~\bibnamefont {Arbouet}},\ }\bibfield  {title} {\bibinfo
  {title} {Polarizabilities of complex individual dielectric or plasmonic
  nanostructures},\ }\href {https://doi.org/10.1103/physrevb.101.235418}
  {\bibfield  {journal} {\bibinfo  {journal} {Phys. Rev. B}\ }\textbf {\bibinfo
  {volume} {101}},\ \bibinfo {pages} {235418} (\bibinfo {year}
  {2020})}\BibitemShut {NoStop}%
\bibitem [{\citenamefont {Dresselhaus}\ \emph {et~al.}(2008)\citenamefont
  {Dresselhaus}, \citenamefont {Dresselhaus},\ and\ \citenamefont
  {Jorio}}]{Dresselhaus}%
  \BibitemOpen
  \bibfield  {author} {\bibinfo {author} {\bibfnamefont {M.~S.}\ \bibnamefont
  {Dresselhaus}}, \bibinfo {author} {\bibfnamefont {G.}~\bibnamefont
  {Dresselhaus}},\ and\ \bibinfo {author} {\bibfnamefont {A.}~\bibnamefont
  {Jorio}},\ }\href@noop {} {\emph {\bibinfo {title} {{Group Theory.
  Application to the Physics of Condensed Matter}}}}\ (\bibinfo  {publisher}
  {Springer, Berlin},\ \bibinfo {year} {2008})\BibitemShut {NoStop}%
\bibitem [{\citenamefont {Belov}\ and\ \citenamefont
  {Simovski}(2005)}]{Belov2005}%
  \BibitemOpen
  \bibfield  {author} {\bibinfo {author} {\bibfnamefont {P.~A.}\ \bibnamefont
  {Belov}}\ and\ \bibinfo {author} {\bibfnamefont {C.~R.}\ \bibnamefont
  {Simovski}},\ }\bibfield  {title} {\bibinfo {title} {Homogenization of
  electromagnetic crystals formed by uniaxial resonant scatterers},\ }\href
  {https://doi.org/10.1103/physreve.72.026615} {\bibfield  {journal} {\bibinfo
  {journal} {Phys. Rev. E}\ }\textbf {\bibinfo {volume} {72}},\ \bibinfo
  {pages} {026615} (\bibinfo {year} {2005})}\BibitemShut {NoStop}%
\end{thebibliography}%

\end{document}